\documentclass{article}

\PassOptionsToPackage{numbers, sort&compress}{natbib}
\PassOptionsToPackage{table}{xcolor}

\usepackage[preprint]{neurips_2026}

\usepackage[utf8]{inputenc}      
\usepackage[T1]{fontenc}         
\usepackage{hyperref}            
\usepackage{url}                 
\usepackage{booktabs}            
\usepackage{amsfonts}            
\usepackage{nicefrac}            
\usepackage{microtype}           
\usepackage{amsmath}             
\usepackage{amssymb}
\usepackage{graphicx}
\usepackage{multirow}
\usepackage{array}
\usepackage{makecell}
\usepackage{caption}
\usepackage{placeins}
\usepackage{booktabs}
\usepackage{multirow}
\usepackage[table]{xcolor}
\usepackage{titlesec}
\titlespacing*{\paragraph}{0pt}{0.6ex plus 0.2ex}{0.6em}
\hypersetup{
  hidelinks,
  pdftitle={RxnOptBench: Benchmarking LLMs for Reaction-Condition Optimization in Organic Methodology},
  pdfauthor={Lingli Ge, Yubin Wang, Junyuan Gao, Jiahe Song, Jiaxing Sun, Boyu Zhu, Haote Yang, Jingchao Wang, Lixin Ma, Jiang Wu, Yuqiang Li, Conghui He},
  pdfkeywords={AI for Science, reaction condition optimization, multi-objective evaluation, Organic Methodology}
}

\usepackage[most]{tcolorbox}
\tcbuselibrary{listings, breakable, skins}
\definecolor{promptframe}{HTML}{4A6A8A}
\definecolor{promptbg}{HTML}{F7F9FC}
\definecolor{promptkw}{HTML}{2B5D8B}
\lstdefinelanguage{rxnprompt}{
  morekeywords={Reactants, Reactant, Target, Product,
                Reaction, Fixed, Conditions, Constant, Reference,
                Candidate, Values, Options, Option, Output, Format,
                Reagents, Solvents, Catalysts, SMILES, Important, yield},
  morecomment=[l]{\#\#\#},
  sensitive=true,
}
\newtcblisting{promptbox}[2][]{%
  enhanced, breakable,
  colback=promptbg, colframe=promptframe,
  colbacktitle=promptframe, coltitle=white,
  fonttitle=\bfseries\small,
  attach boxed title to top left={xshift=8pt, yshift=-7pt},
  boxed title style={sharp corners, rounded corners=northwest,
                     colframe=promptframe, size=fbox},
  arc=2pt, outer arc=2pt,
  left=6pt, right=6pt, top=10pt, bottom=4pt,
  listing only,
  listing options={
    language=rxnprompt,
    basicstyle=\ttfamily\scriptsize,
    keywordstyle=\color{promptkw}\bfseries,
    commentstyle=\color{promptframe}\bfseries,
    breaklines=true,
    breakindent=0pt,
    breakatwhitespace=true,
    columns=fullflexible,
    keepspaces=true,
    upquote=true,
    showstringspaces=false,
    aboveskip=0pt, belowskip=0pt,
  },
  title=#2, #1
}

\title{RxnOptBench: Benchmarking LLMs for Reaction-Condition Optimization in Organic Methodology}

\author{%
  \begin{tabular}{c}
  Lingli Ge$^{1,2,3}$ \quad Yubin Wang$^{3}$ \quad Junyuan Gao$^{3}$ \quad
  Jiahe Song$^{1,3}$ \\
  Jiaxing Sun$^{3}$ \quad Boyu Zhu$^{3}$ \quad Haote Yang$^{3}$ \quad
  Jingchao Wang$^{3}$ \\
  Lixin Ma$^{3}$ \quad Jiang Wu$^{3}$\textsuperscript{*} \quad Yuqiang Li$^{2}$\textsuperscript{*} \quad
  Conghui He$^{3}$\textsuperscript{*} \\[0.5em]
  {\normalfont\small $^{1}$Shanghai Jiao Tong University, Shanghai, China} \\
  {\normalfont\small $^{2}$Shanghai Innovation Institute, Shanghai, China} \\
  {\normalfont\small $^{3}$Shanghai Artificial Intelligence Laboratory, Shanghai, China}
  \end{tabular}
}

\begin{document}

\maketitle
\begingroup
\renewcommand{\thefootnote}{*}
\footnotetext{Corresponding authors.}
\endgroup

\begin{abstract}
Chemical reaction-condition optimization --- choosing the catalyst, ligand, solvent, reagent, temperature, time, and atmosphere that jointly maximize yield and stereoselectivity --- is a central, judgement-laden subtask of organic methodology research that large language models are increasingly expected to support. Yet existing chemistry benchmarks evaluate reaction-class labelling, retrosynthesis, or SMILES manipulation, and do not ask models to read a real condition-screening table and pick the best set. We introduce \textbf{RxnOptBench}, a benchmark whose every option and precedent is a real wet-lab entry mined from the optimization tables of organic-methodology papers published in 2025, graded by a continuous relative score derived from a declared headline utility that combines reported yield with enantiomeric excess (ee), diastereomeric ratio (dr), and regioisomeric ratio (rr), and equipped with a paired precedents-vs-no-precedents design that isolates in-context use of literature evidence from parametric memorization. Across nine frontier LLMs and three Chemistry LLMs, even the best models leave substantial headroom: chemistry-specialized models fall to the random-baseline floor on multi-axis selection, while open-weight models have closed most of the gap to proprietary frontier models. We release the final human-reviewed benchmark test set and evaluation code.
\end{abstract}

\begin{figure}[t]
  \centering
  \includegraphics[width=\linewidth]{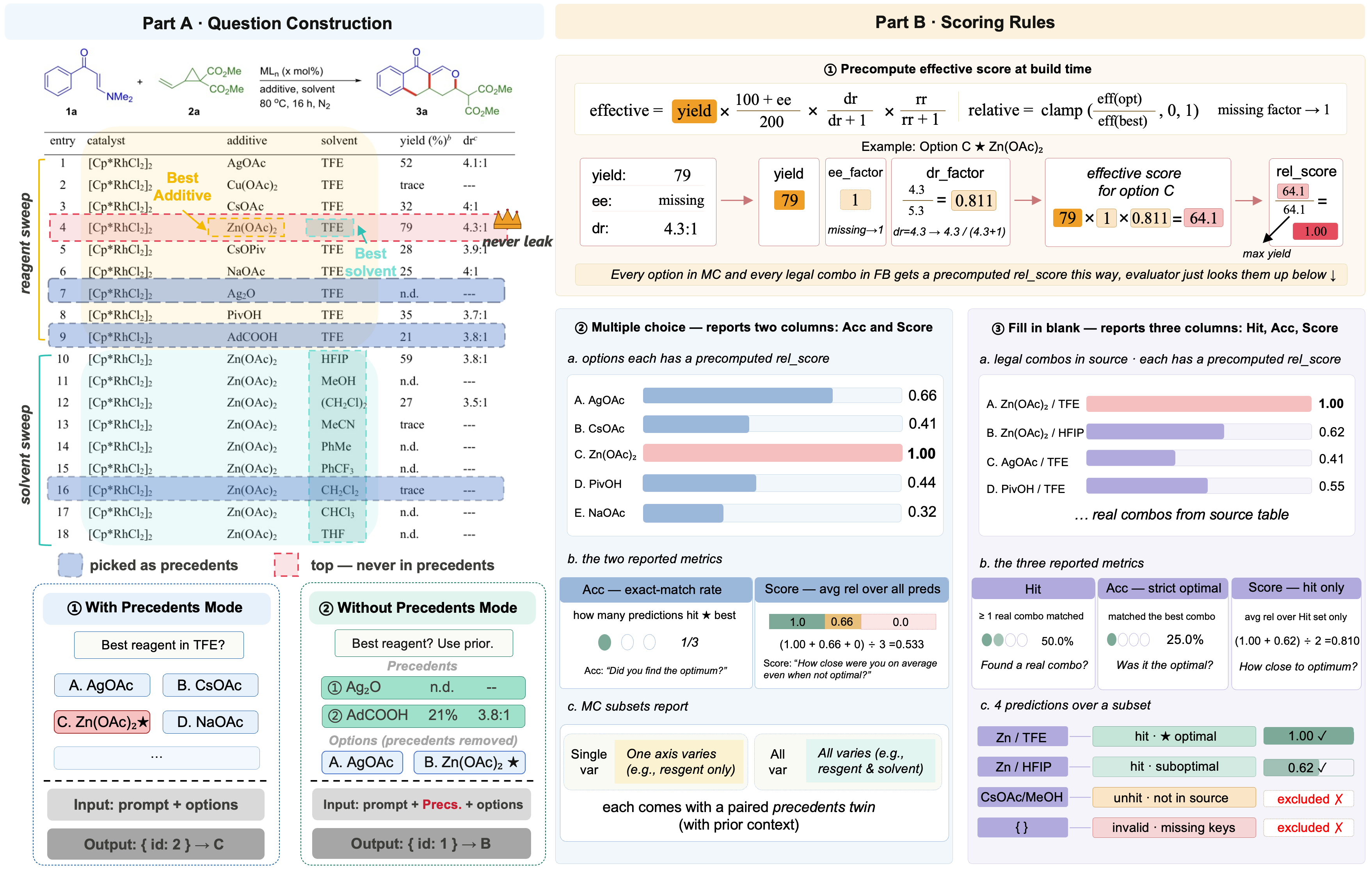}
  \caption{\textbf{RxnOptBench overview.} (A) Question Construction: each condition-screening table yields MC/FIB items in matched \emph{precedents} (prior rows shown, top withheld) and \emph{no-precedents} (no prior rows) modes. (B) Scoring Rules: every observed row carries a precomputed relative score fusing yield with ee/dr/rr. MC reports Accuracy and Score; FIB separately reports observed-row Hit, Strict Accuracy, and Score conditioned on Hit.}
  \label{fig:overview}
\end{figure}

\section{Introduction}

Over the past two decades, organic synthesis methodology has remained a central driver of progress in chemistry, reflected by multiple Nobel Prizes recognizing advances such as asymmetric catalysis, olefin metathesis, cross-coupling, and click chemistry. These breakthroughs were not defined solely by the discovery of new bond-forming transformations, but by the development of reaction conditions that rendered them broadly practical and reproducible across diverse substrates. Catalysts, ligands, solvents, additives, and reaction environments collectively determine whether a transformation can achieve the efficiency and selectivity required for applications in pharmaceutical, agrochemical, and materials synthesis. Contemporary methodology research continues to expand this landscape, with new catalytic systems and activation modes reported each year across thermal, photochemical, electrochemical, and multi-catalytic regimes.

A central challenge underlying this process is \emph{reaction condition optimization}: identifying the combination of catalyst, ligand, solvent, reagent, temperature, and other controllable variables that maximizes yield and stereoselectivity for a target transformation. Existing computational approaches address this problem only partially. Bayesian optimization and high-throughput experimentation frameworks \citep{shields2021bayesian,ahneman2018predicting,gao2018using} can efficiently search within predefined experimental spaces, but they typically assume a fixed reaction class and a closed candidate set. Such approaches therefore offer limited support when reasoning about newly reported transformations or unfamiliar catalytic systems. In parallel, large language models (LLMs) have recently emerged as general-purpose reasoning systems for chemistry and now constitute the core component of many autonomous or semi-autonomous chemistry agents \citep{boiko2023autonomous,bran2024chemcrow}. These systems derive their capabilities from large-scale exposure to the chemical literature, suggesting the possibility of leveraging accumulated methodological knowledge for reaction-condition recommendation. However, whether current LLMs can reliably perform experimentally meaningful condition selection remains largely unexamined.

Existing chemistry benchmarks provide little insight into this capability. Widely used reaction datasets, including USPTO-50k and related variants \citep{lowe2012extraction,schneider2016whats,liu2017retrosynthetic}, primarily support retrosynthesis and reaction prediction tasks. Larger reaction corpora such as Pistachio \citep{nextmove2017pistachio} and the Open Reaction Database \citep{kearnes2021open} record successful experimental outcomes, but generally preserve reactions as isolated entries rather than as matched sets of tested alternatives with directly comparable outcomes. More recent chemistry benchmarks --- including ChemLLMBench \citep{guo2023chemllmbench}, ChemBench \citep{mirza2025chembench}, SciBench \citep{wang2024scibench}, and ChemVLM \citep{li2025chemvlm} --- evaluate molecular property prediction, reaction understanding, SMILES manipulation, or general scientific reasoning. Collectively, these benchmarks measure whether models can retrieve or manipulate chemical knowledge, but not whether they can make the comparative decisions required during experimental optimization.

Organic-methodology publications commonly include \emph{condition-screening tables}, which disclose a curated subset of experimentally tested conditions supporting a reported transformation. This curation defines which local decision spaces are visible; among the disclosed rows, the benchmark uses the measured outcomes and their ordering directly as labels. Although these tables are not complete optimization histories, they preserve matched comparisons for the same reaction while varying catalyst, ligand, solvent, base, additive, temperature, or related factors. They therefore support an experimentally grounded conditional task: selecting the best reported condition among the candidates disclosed in that table (Figure~\ref{fig:overview}A). Without additional wet-lab validation, unreported combinations have no defensible outcome label and are treated as outside measured support, not as failures or negative examples; the best disclosed row is not interpreted as a global optimum over chemical space.

Motivated by this observation, we introduce \textbf{RxnOptBench}, a benchmark designed to evaluate LLMs on reaction-condition recommendation. We transform condition-screening tables into multiple-choice and fill-in-the-blank tasks that require models to identify optimal or near-optimal reaction conditions. Following a four-stage extraction pipeline with full human verification (Section~\ref{sec:pipeline}), RxnOptBench comprises \textbf{4{,}773 questions across eight task sub-types} (3{,}911 multiple-choice and 862 fill-in-the-blank questions), derived from \textbf{552 screening tables spanning 232 organic-methodology papers published in 2025 across 16 journals}. This recent sampling frame deliberately targets contemporary, potentially unfamiliar condition-selection problems rather than canonical-recipe recall; recency reduces but does not eliminate exact source exposure. Answers receive partial credit through the benchmark's declared headline aggregate utility, which combines reported yield and stereoselectivity but is not intended to encode every chemical objective or chemist preference. RxnOptBench evaluates both cold-start and precedent-assisted reasoning and separately probes single-factor versus full-condition selection.

We evaluate nine frontier general-purpose LLMs together with three chemistry-specialized LLMs on RxnOptBench. The results reveal four main findings. First, substantial performance headroom remains even for the strongest proprietary models, indicating that reaction-condition recommendation is far from a solved problem. Second, tasks requiring global reasoning over complete condition sets are consistently more challenging than single-factor selection. Third, chemistry-specialized models do not consistently outperform frontier general-purpose LLMs, particularly on the most difficult task settings. Finally, providing experimentally related precedents yields measurable improvements, suggesting that LLMs can partially leverage literature context for condition selection rather than relying solely on memorized associations.

\begin{figure}[t]
\centering
\includegraphics[width=\linewidth]{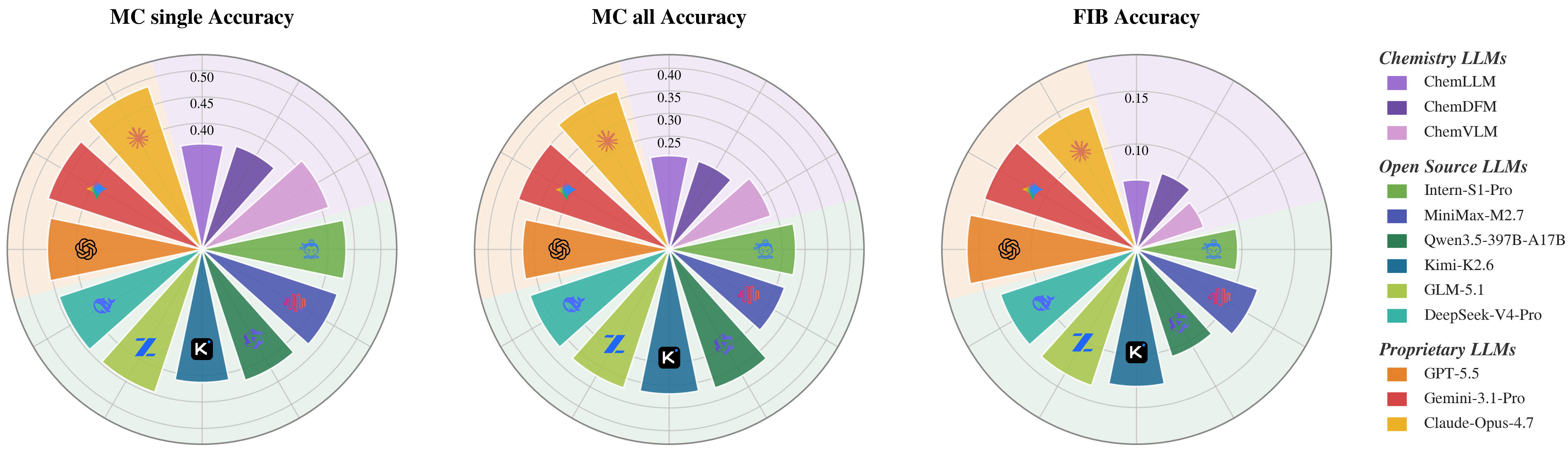}
\caption{\textbf{Headline accuracy on the no-precedents subsets.} Per-model Accuracy on \texttt{MC\_single} ($n=1{,}039$), \texttt{MC\_all} ($n=552$), and FIB ($n=552$); sectors group models by tier. FIB uses Strict Accuracy over submitted predictions. Numbers in Table~\ref{tab:all_heatmap}; precedents comparisons use the remaining matched subsets.}
\label{fig:fullset_rose}
\end{figure}

\section{Related Work}

\paragraph{Reaction datasets and condition-prediction baselines.}
USPTO-50k and its variants \citep{lowe2012extraction,schneider2016whats,liu2017retrosynthetic} and Pistachio \citep{nextmove2017pistachio} have driven a decade of work on reaction-class labelling and retrosynthesis, but record only individual reactions, stripped from the screening context in which they were chosen. The Open Reaction Database (ORD) \citep{kearnes2021open} provides a structured schema for sharing reaction records, and ORDerly \citep{wigh2024orderly} cleans ORD into ML-ready splits, including the first ORD-derived condition-prediction benchmark. Coley-style predictors \citep{gao2018using}, multilabel cross-coupling models \citep{maser2021multilabel}, sequence-to-sequence models that infer full experimental procedures from reaction SMILES \citep{vaucher2021inferring}, and Bayesian-optimization workflows \citep{shields2021bayesian,ahneman2018predicting} all output condition vectors directly, but assume a fixed reaction template or rely on dense HTE coverage that real methodology tables rarely provide. None, however, frames the task as a chemist does at the bench: choosing the empirically best entry from a small table of pre-screened candidates.

\paragraph{Chemistry LLMs and benchmarks.} Existing benchmarks---ChemLLMBench \citep{guo2023chemllmbench}, ChemBench \citep{mirza2025chembench}, Mol-Instructions \citep{fang2024molinstructions}---evaluate LLMs on property prediction, name-reaction trivia, and SMILES manipulation. Domain models like ChemLLM \citep{zhang2024chemllm} and ChemDFM \citep{zhao2024chemdfm} are pretrained on chemistry corpora, while ChemVLM \citep{li2025chemvlm} extends to the multimodal regime. Broader suites \citep{wang2024scibench,sun2024scieval,lu2022scienceqa} probe curriculum-level reasoning, and ChemTable \citep{zhou2025chemtable} and MMSci-Eval \citep{yang2025mmsci} target scientific tables. A parallel line targets condition or yield prediction: MM-RCR \citep{zhang2024chemmarc}, Chemma \citep{zhang2025large}, and SynAsk \citep{zhang2024synask} extend the Coley-style line \citep{gao2018using} to LLMs, but are trained on USPTO corpora or single-template HTE screens \citep{haywood2024when}. None separate single-axis from whole-condition reasoning, ground options in real screening-table entries, or provide continuous partial credit over reported yield and selectivity. Agentic systems such as ChemCrow \citep{bran2024chemcrow} and Coscientist \citep{boiko2023autonomous}, surveyed by \citet{ramos2025review}, sit one level higher and presuppose the condition-decision capability we benchmark.

\paragraph{Condition recommendation methods and benchmarks.} We group prior automatic condition-recommendation methods into two regimes. Global models train on patent- or literature-scale corpora, including the USPTO-Condition and Reaxys-TotalSyn-Condition benchmarks released with Parrot \citep{wang2023parrot}, the USPTO reagent corpus used by \citet{andronov2023reagent}, and the USPTO-500MT condition split \citep{lu2022unified} adopted by TextReact \citep{qian2023textreact} and MM-RCR \citep{zhang2024chemmarc}, with ChemMAS \citep{yang2025chemmas} further scaling to 544k proprietary reactions. Local models target a single reaction family with HTE data: Bayesian optimization \citep{shields2021bayesian,ahneman2018predicting} has been extended to multi-objective yield-ee tuning \citealp{garridotorres2022edboplus}, bandit-style identification of general conditions \citep{wang2024banditgeneral}, and highly parallel HTE-ML campaigns \citep{sin2025minerva}. All of these produce a condition vector rather than selecting the best entry of candidates, sitting orthogonal to RxnOptBench.

\section{RxnOptBench}

\subsection{Source}
\label{sec:source}

We mine condition-screening tables from organic-methodology papers first
published in 2025 across 16 prespecified venues: \emph{ACS Catal.},
\emph{ACS Cent.\ Sci.}, \emph{Angew.\ Chem.\ Int.\ Ed.}, \emph{Chem},
\emph{Chem Catalysis}, \emph{Chem.\ Sci.}, \emph{Green Chem.}, \emph{JACS},
\emph{JACS Au}, \emph{Nature}, \emph{Nat.\ Catal.}, \emph{Nat.\ Chem.},
\emph{Nat.\ Commun.}, \emph{Nat.\ Synth.}, \emph{Org.\ Lett.}, and
\emph{Sci.\ Adv.} Following the revised venue categorization, JACS and
\emph{Angew.\ Chem.\ Int.\ Ed.} are grouped as chemistry-discipline
flagships, whereas \emph{ACS Cent.\ Sci.} and \emph{Nat.\ Commun.} are
grouped as broad-scope flagships. This presentation-level regrouping does not
change venue inclusion or sampling. Both the main text and the full Supporting
Information are fed to the pipeline.

This one-year frame is an intended evaluation distribution rather than an
estimate of historical publication prevalence. It emphasizes contemporary,
recently disclosed transformations, catalytic systems, and condition regimes,
thereby probing transfer to recent chemical decision problems. We do not
assume that every transformation constitutes a new reaction class.

After passing the extraction pipeline (Section~\ref{sec:pipeline}) and full manual expert verification, the resulting corpus statistics are deferred to Section~\ref{sec:pipeline}. The exclusion of patent-mined and aggregator sources, whose row-level outcomes are routinely truncated, re-normalized, or re-typeset, keeps every source artifact directly auditable against the native PDF. To avoid inflating chemical-space counts by the multiple questions derived from one source table, we analyze composition over the 552 parent screening tables. They span 20 primary transformation classes, 26 condition fields, and 341 unique canonical products containing 232 non-empty Bemis--Murcko scaffolds; the reaction, condition, and product distributions are reported in Appendix~\ref{app:coverage}.

\subsection{Extraction Pipeline}
\label{sec:pipeline}

\begin{figure*}[t]
  \centering
  \captionsetup{skip=3pt}
  \includegraphics[width=0.96\textwidth]{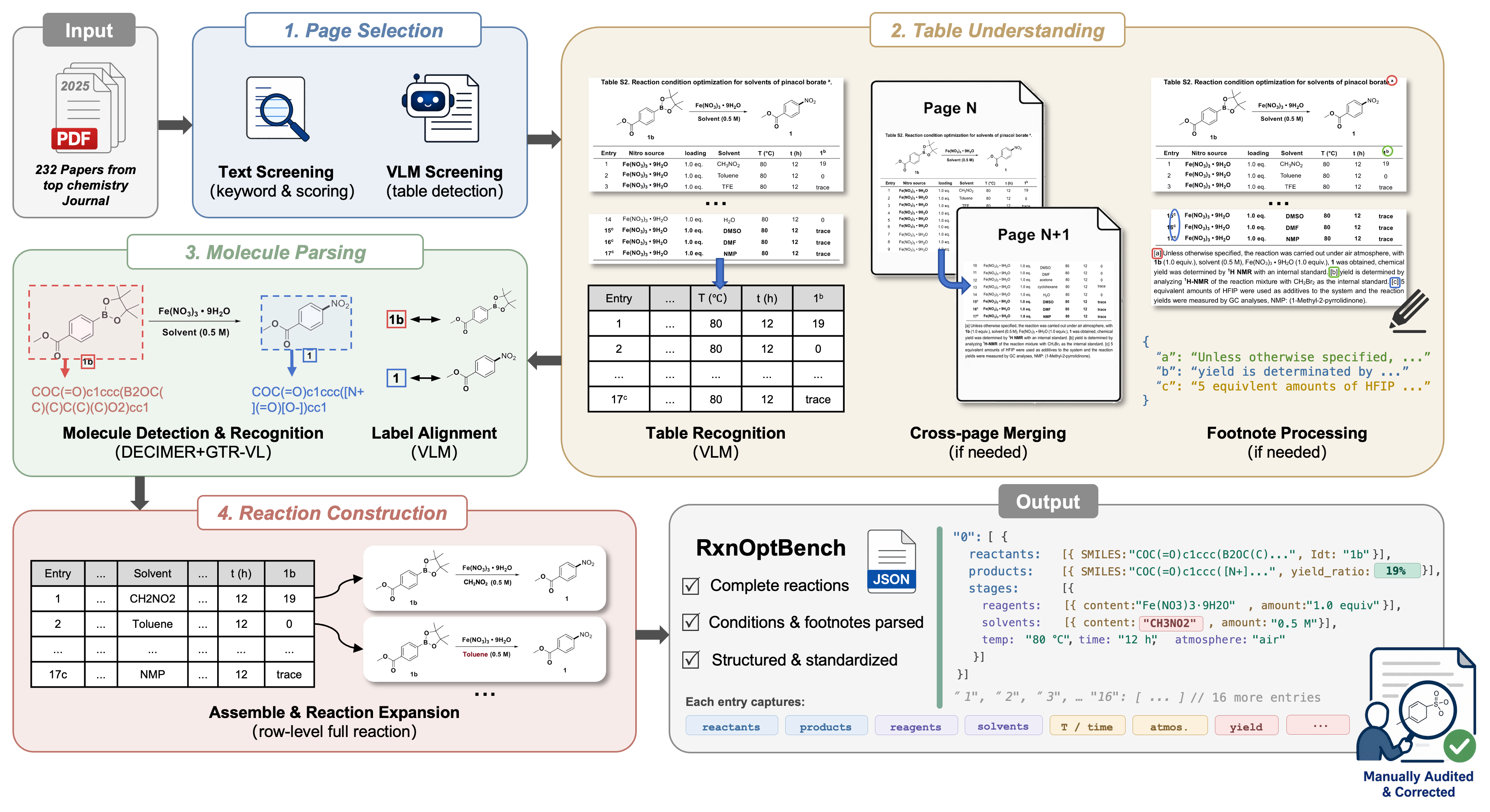}
  \caption{\textbf{Four-stage semi-automated extraction pipeline.}
  PDF pages are screened, parsed, converted into row-level reaction records,
  and manually audited before question construction. The GTR-VL label
  in Stage~3 denotes the MinerU.Chem recognizer; details in
  \S\ref{sec:pipeline}.}
  \label{fig:pipeline}
\end{figure*}

A four-stage semi-automated pipeline (Figure~\ref{fig:pipeline}) converts the 232 source PDFs into \textbf{552 retained tables} (of 911 candidates) and \textbf{5{,}451 row-level reaction records}, each carrying resolved substrate/product/catalyst/ligand SMILES, a canonical condition string, measured outcomes (yield, ee, d.r.), and a back-pointer to paper/page/row. Visual stages run on 600\,DPI page PNGs; each stage pairs a strict-format VLM call with a deterministic check.

\paragraph{Page Selection \& Table Understanding.}
A regex-then-VLM cascade shortlists pages whose captions or body text mention optimization/screening and separates true screening tables from substrate-scope, mechanistic, and characterization tables; a second VLM emits structured \texttt{\{caption, footnotes, columns, rows\}} per table, with cross-page tables reconstructed from neighbours. All VLM calls in the extraction pipeline other than molecular-structure recognition use Kimi-K2.6. These stages are described for methodological transparency, but their prompts, parsing code, and runtime implementation are not part of the public release.

\paragraph{Molecule Parsing.}
DECIMER-Segmentation~\citep{rajan2021decimerseg} (Mask~R-CNN~\citep{he2017maskrcnn}) localizes molecule drawings. We use the GTR-VL-based molecular-structure recognition module of MinerU.Chem~\citep{yang2026mineruchem}, accessed through an API, to recognize each crop; model-specific post-processing converts its output into structures validated by RDKit~\citep{landrum2024rdkit}. A VLM then binds abstract labels (\texttt{1a}, \texttt{L1}, \texttt{Pd-1}) to these structures via SMILES lookup --- it never writes structures. The MinerU.Chem recognition service, our model-specific post-processing, and the broader extraction and benchmark-construction code are not redistributed with RxnOptBench. The public release is limited to the final human-reviewed benchmark test set and evaluation code; it supports consistent scoring of model predictions but not reconstruction from raw PDFs.

\paragraph{Reaction Construction \& Final Expert Review.}
Per-row constrained VLM calls type each cell (yield, ee, dr, solvent, temperature, \dots), populate substrate/product/catalyst/ligand slots, and expand footnote/arrow annotations into a canonical condition string; text-only reagents are resolved via curated dictionary $\to$ local cache $\to$ PubChem~\citep{kim2025pubchem}. The checks within Stages~1--4 are deterministic rather than human reviews. After Stage~4 has assembled complete table- and row-level records, every extracted table is compared with the source PDF by four master's-level and two doctoral-level paid chemistry interns based in mainland China, who edit, accept, or reject it; only records passing this final gate enter question construction. The automatic stages support batched inputs, concurrent calls, and resumable intermediate outputs; expert verification is the deliberate final quality gate and the main curation bottleneck, so future growth proceeds through versioned reviewed releases.

\subsection{Question Construction}
\label{sec:question_construction}

A retained table reaches question construction only if it contains $\geq\!2$ rows whose normalized condition strings differ on at least one axis (normalization: Unicode NFKC, sub-/super-script folding, whitespace and case folding); single-row tables and tables collapsing to a single condition string are excluded. A condition category is varying if its normalized value differs across $\geq\!2$ entries; table-level constants enter the prompt as context, while varying axes define the candidate space. RxnOptBench then defines two response types and eight evaluated subsets (Table~\ref{tab:subtypes}).

\begin{table}[h]
\centering
\caption{The eight sub-types of RxnOptBench.}
\label{tab:subtypes}
\footnotesize
\setlength{\tabcolsep}{4pt}
\begin{tabular}{l l r p{6cm}}
\toprule
\textbf{Type} & \textbf{Sub-type} & \textbf{\#Q} & \textbf{Description} \\
\midrule
MC  & \texttt{single\_var}             & 1{,}039 & One target axis; all other varying axes locked. \\
MC  & \texttt{all\_var}                &   552   & Choose one observed joint-condition row. \\
MC  & \texttt{single\_var\_precs}        &   850   & \texttt{single\_var} with precedents. \\
MC  & \texttt{all\_var\_precs}           &   310   & \texttt{all\_var} with precedents. \\
MC  & \texttt{single\_var\_precs\_paired}  &   850   & Twin of \texttt{single\_var\_precs} without precedents. \\
MC  & \texttt{all\_var\_precs\_paired}     &   310   & Twin of \texttt{all\_var\_precs} without precedents. \\
FIB & \texttt{all\_blank}                &   552   & One index per axis; observed joint tuple required. \\
FIB & \texttt{all\_blank\_precs}         &   310   & \texttt{all\_blank} with precedents. \\
\midrule
\multicolumn{2}{l}{\textbf{Total}} & \textbf{4{,}773} & 3{,}911 MC + 862 FIB. \\
\bottomrule
\end{tabular}
\end{table}

Both MC families are \emph{single-answer}: \texttt{MC\_single} locks all but
one varying axis to a real fixed tuple and scores within that local group,
whereas each \texttt{MC\_all} option is a complete observed row. FIB returns
one index per varying key; independently observed values can recombine into an
untested tuple, so Hit requires the complete tuple to match an observed row.
Normalized duplicates retain the highest-scoring row and all maximum ties are
accepted as gold (formal construction in Appendix~\ref{app:subtypes}).

Two construction details not visible in the table: precedents are
condition-disjoint from every option (their yield-band sampling and matched
no-precedents pairing rule are deferred to \S\ref{sec:refctrl}), and all
formal evaluations use shuffled candidate order. MC option bundles are
shuffled by a question-ID-derived seed; each FIB candidate list is shuffled
independently by question and condition key using a stable hash and seed
\texttt{20260504}. The same resulting question version and index space are
used for all 12 models, and the public test-set records store each exact
permutation.

\subsection{Metrics}
\label{sec:metrics}

Each entry carries the benchmark's headline aggregate utility: a monotone,
bounded summary combining reported yield with available selectivity. It
provides one reproducible scalar ranking but is not intended to encode every
chemical objective or chemist preference,
\begin{equation}
\text{effective}(s) = \text{yield}\cdot f_{\mathrm{enantio}}\cdot
f_{dr}\cdot f_{rr},
\end{equation}
where
\begin{equation}
f_{\mathrm{enantio}} =
\begin{cases}
(100+|ee|)/200, & ee\ \text{reported},\\
\rho_{er}/(\rho_{er}+1), & er\ \text{reported but }ee\ \text{absent},\\
1, & \text{otherwise},
\end{cases}
\quad
f_{dr}=\frac{\rho_{dr}}{\rho_{dr}+1},\quad
f_{rr}=\frac{\rho_{rr}}{\rho_{rr}+1}.
\end{equation}
Here $ee$ denotes enantiomeric excess, while $er$, $dr$, and $rr$ denote
enantiomeric, diastereomeric, and regioisomeric ratios, respectively. Each
factor is the major-product fraction on its selectivity axis: major enantiomer
for $ee/er$, major diastereomer for $dr$, and major regioisomer for $rr$. For
each ratio metric, $\rho$ is the normalized major:minor ratio (thus either
reported order, e.g.\ $20{:}1$ or $1{:}20$, maps to $\rho=20$). An unreported
selectivity axis contributes the neutral factor $1$; this omits that axis from
the utility rather than assuming perfect selectivity.

For question $q$ with scorable set $\mathcal O_q$, define
\begin{equation}
r_q(o)=\mathrm{clamp}\!\left(\text{effective}(o)/U_q^\star,0,1\right),
\qquad U_q^\star=\max_{o'\in\mathcal O_q}\text{effective}(o'),
\end{equation}
with equal zero numerator/denominator defined as $1$. Here $\mathcal O_q$ is
global for \texttt{MC\_all}/FIB and local for \texttt{MC\_single}. For
$A_q=\{i:r_q(o_{q,i})=1\}$ and MC prediction $\hat i_q$, over $N$ submissions
\begin{equation}
\mathrm{Acc}_{\mathrm{MC}}=\frac{1}{N}\sum_q
\mathbf 1[\hat i_q\in A_q],\qquad
\mathrm{Score}_{\mathrm{MC}}=\frac{1}{N}\sum_q r_q(o_{q,\hat i_q}),
\end{equation}
where an invalid index contributes zero to both.

For FIB, let $V_q$ mark a valid in-range mapping $\hat{\mathbf x}_q$, and define
$H_q=\mathbf 1[V_q=1\land\hat{\mathbf x}_q\in\mathcal L_q]$ and
$B_q=H_q\,\mathbf 1[r_q(\hat{\mathbf x}_q)=1]$. The reported metrics are
\begin{equation}
\mathrm{Hit}=\frac{\sum_q H_q}{N},\qquad
\mathrm{StrictAcc}=\frac{\sum_q B_q}{N},\qquad
\mathrm{Score}\mid\mathrm{Hit}=
\frac{\sum_{q:H_q=1}r_q(\hat{\mathbf x}_q)}{\sum_q H_q}.
\end{equation}
Thus $\mathrm{StrictAcc}=\mathrm{Hit}\Pr(B_q=1\mid H_q=1)$. Invalid and
non-Hit outputs fail Hit/Strict but lack measured outcomes and are excluded
from Score$\,\mid\,$Hit; non-Hit does not mean chemically invalid.
Appendix~\ref{app:robustness} gives score-sensitivity analyses.
Table~\ref{tab:all_heatmap} reports a matched uniform-random floor
($\sim\!0.54$--$0.67$), against which Scores should be read.

\subsection{Precedents vs.\ No-precedents: Paired Design}
\label{sec:refctrl}

Each \texttt{*\_precs} question is paired with a \texttt{*\_precs\_paired} twin sharing identical option space, answer key, and scoring; they differ only in whether 2--5 precedent experiments are appended. Precedents are drawn from the lower-to-mid yield band, the optimum is withheld, and (FIB) precedent combinations are removed from legal answers --- so $\Delta r = r_{\text{precs}}\!-\!r_{\text{noprec}}$ isolates trend extrapolation, free of difficulty/option/scoring confounds.

\subsection{Models and Inference Protocol}
\label{sec:models}

We evaluate nine frontier LLMs --- six Open Source LLMs (DeepSeek-V4-Pro \citep{deepseek2025v4}, Kimi-K2.6 \citep{moonshot2025kimik26}, GLM-5.1 \citep{zhipu2025glm51}, Qwen3.5-397B-A17B \citep{qwen2025team}, Intern-S1-Pro \citep{zou2026intern}, MiniMax-M2.7 \citep{minimax2026m27}) and three Proprietary LLMs (Claude-Opus-4.7 \citep{anthropic2025claudeopus}, GPT-5.5 \citep{openai2025gpt55}, Gemini-3.1 Pro thinking variant \citep{google2025gemini31}) --- plus three Chemistry LLMs (ChemLLM \citep{zhang2024chemllm}, ChemVLM \citep{li2025chemvlm}, ChemDFM \citep{zhao2024chemdfm}). All models share one prompt template per question type. Sampling is disabled (\texttt{temperature=0} where available, otherwise provider greedy/non-sampling mode). The public evaluator deterministically scores supplied predictions against the fixed, human-reviewed test set; hosted backends need not reproduce generations bit-for-bit (Appendix~\ref{app:robustness}).

\section{Empirical Results}
\label{sec:findings}

\begin{table}[t]
\centering
\definecolor{heatlow}{RGB}{255,255,212}
\definecolor{heatmid}{RGB}{254,196,79}
\definecolor{heathigh}{RGB}{236,112,20}
\definecolor{rowChem}{HTML}{FFE8D0}
\definecolor{rowOpen}{HTML}{DDEFE0}
\definecolor{rowClosed}{HTML}{D8E2F4}
\setlength{\tabcolsep}{3pt}
\renewcommand{\arraystretch}{1.05}
\newcolumntype{C}{>{\centering\arraybackslash}p{0.85cm}}
\newcolumntype{T}{>{\centering\arraybackslash}p{1.25cm}}
\scriptsize
\caption{No-precedents results (\texttt{MC\_single}: 1,039; \texttt{MC\_all}/FIB: 552 each). FIB Strict uses all submissions; Score is conditional on Hit. Cells are heat-mapped per column; models are ordered weakest$\to$strongest by mean Score within tier.}
\begin{tabular}{T l*{7}{C}}
\toprule
 & \multirow{2}{*}{\textbf{Model}}
 & \multicolumn{2}{c}{\textbf{MC\_single}}
 & \multicolumn{2}{c}{\textbf{MC\_all}}
 & \multicolumn{3}{c}{\textbf{FIB}} \\
\cmidrule(lr){3-4}\cmidrule(lr){5-6}\cmidrule(lr){7-9}
 & & Acc. & Score & Acc. & Score & Hit & Strict & Score$|$Hit \\
\midrule
\textit{Baseline} & \textit{Uniform random} & 36.5\% & 0.669 & 17.8\% & 0.540 & 28.1\% & 5.9\% & 0.540 \\
\midrule
\multirow{3}{*}{\makecell[c]{\textbf{Chemistry}\\\textbf{LLMs}}}
& ChemLLM  & \cellcolor{heatlow}36.1\%            & \cellcolor{heatlow}0.666            & \cellcolor{heatmid!2!heatlow}20.7\%  & \cellcolor{heatlow}0.545            & \cellcolor{heatlow}25.2\%             & \cellcolor{heatlow}6.6\%             & \cellcolor{heatlow}0.554 \\
& ChemDFM  & \cellcolor{heatmid!10!heatlow}36.7\% & \cellcolor{heatmid!8!heatlow}0.670  & \cellcolor{heatlow}20.5\%            & \cellcolor{heatmid!2!heatlow}0.547  & \cellcolor{heatmid!68!heatlow}33.5\%  & \cellcolor{heatmid!21!heatlow}7.6\%  & \cellcolor{heatmid!4!heatlow}0.557 \\
& ChemVLM  & \cellcolor{heatmid!85!heatlow}41.4\% & \cellcolor{heatmid!94!heatlow}0.712 & \cellcolor{heatmid!38!heatlow}23.6\% & \cellcolor{heatmid!63!heatlow}0.602 & \cellcolor{heatmid!20!heatlow}27.7\%  & \cellcolor{heatmid!2!heatlow}6.7\%   & \cellcolor{heatmid!77!heatlow}0.610 \\
\midrule
\multirow{6}{*}{\makecell[c]{\textbf{Open Source}\\\textbf{LLMs}}}
& Intern-S1-Pro     & \cellcolor{heathigh!17!heatmid}43.4\% & \cellcolor{heathigh!39!heatmid}0.734 & \cellcolor{heatmid!91!heatlow}27.9\%  & \cellcolor{heathigh!10!heatmid}0.644 & \cellcolor{heathigh!34!heatmid}41.7\% & \cellcolor{heatmid!63!heatlow}9.6\%  & \cellcolor{heatmid!30!heatlow}0.576 \\
& Qwen3.5-397B-A17B & \cellcolor{heatmid!98!heatlow}42.2\% & \cellcolor{heatmid!92!heatlow}0.711 & \cellcolor{heathigh!44!heatmid}32.2\% & \cellcolor{heathigh!51!heatmid}0.681 & \cellcolor{heathigh!38!heatmid}42.2\% & \cellcolor{heatmid!86!heatlow}10.7\% & \cellcolor{heatmid!84!heatlow}0.615 \\
& Kimi-K2.6         & \cellcolor{heatmid!83!heatlow}41.3\% & \cellcolor{heatmid!84!heatlow}0.707 & \cellcolor{heathigh!40!heatmid}31.9\% & \cellcolor{heathigh!43!heatmid}0.674 & \cellcolor{heathigh!28!heatmid}40.9\% & \cellcolor{heathigh!35!heatmid}13.0\% & \cellcolor{heathigh!14!heatmid}0.637 \\
& GLM-5.1           & \cellcolor{heathigh!36!heatmid}44.6\% & \cellcolor{heathigh!29!heatmid}0.729 & \cellcolor{heathigh!44!heatmid}32.2\% & \cellcolor{heathigh!44!heatmid}0.675 & \cellcolor{heathigh!60!heatmid}44.9\% & \cellcolor{heathigh!47!heatmid}13.6\% & \cellcolor{heathigh!27!heatmid}0.646 \\
& MiniMax-M2.7      & \cellcolor{heathigh!12!heatmid}43.1\% & \cellcolor{heathigh!29!heatmid}0.729 & \cellcolor{heatmid!77!heatlow}26.8\%  & \cellcolor{heathigh!23!heatmid}0.656 & \cellcolor{heathigh!43!heatmid}42.8\% & \cellcolor{heathigh!16!heatmid}12.1\% & \cellcolor{heathigh!54!heatmid}0.666 \\
& DeepSeek-V4-Pro   & \cellcolor{heathigh!36!heatmid}44.6\% & \cellcolor{heathigh!61!heatmid}0.745 & \cellcolor{heathigh!46!heatmid}32.4\% & \cellcolor{heathigh!46!heatmid}0.676 & \cellcolor{heathigh!81!heatmid}47.5\% & \cellcolor{heathigh!47!heatmid}13.6\% & \cellcolor{heathigh!63!heatmid}0.672 \\
\midrule
\multirow{3}{*}{\makecell[c]{\textbf{Proprietary}\\\textbf{LLMs}}}
& GPT-5.5           & \cellcolor{heathigh!49!heatmid}45.4\%          & \cellcolor{heathigh!80!heatmid}0.754          & \cellcolor{heathigh!46!heatmid}32.4\%          & \cellcolor{heathigh!43!heatmid}0.674          & \cellcolor{heathigh!94!heatmid}\underline{49.1\%} & \cellcolor{heathigh}\textbf{16.1\%}   & \cellcolor{heathigh}\textbf{0.699} \\
& Gemini-3.1-Pro    & \cellcolor{heathigh!71!heatmid}\underline{46.8\%} & \cellcolor{heathigh!94!heatmid}\underline{0.760} & \cellcolor{heathigh!79!heatmid}\underline{35.0\%} & \cellcolor{heathigh!82!heatmid}\underline{0.709} & \cellcolor{heathigh}\textbf{49.6\%}   & \cellcolor{heathigh!81!heatmid}\underline{15.0\%} & \cellcolor{heathigh!68!heatmid}0.675 \\
& Claude-Opus-4.7   & \cellcolor{heathigh}\textbf{48.6\%}            & \cellcolor{heathigh}\textbf{0.764}            & \cellcolor{heathigh}\textbf{36.8\%}            & \cellcolor{heathigh}\textbf{0.725}            & \cellcolor{heathigh!56!heatmid}44.4\% & \cellcolor{heathigh!62!heatmid}14.3\% & \cellcolor{heathigh!78!heatmid}\underline{0.683} \\
\bottomrule
\end{tabular}
\label{tab:all_heatmap}
\end{table}


\subsection{Headline Results}
\label{sec:results}

\begin{figure}[t]
\centering
\includegraphics[width=\linewidth]{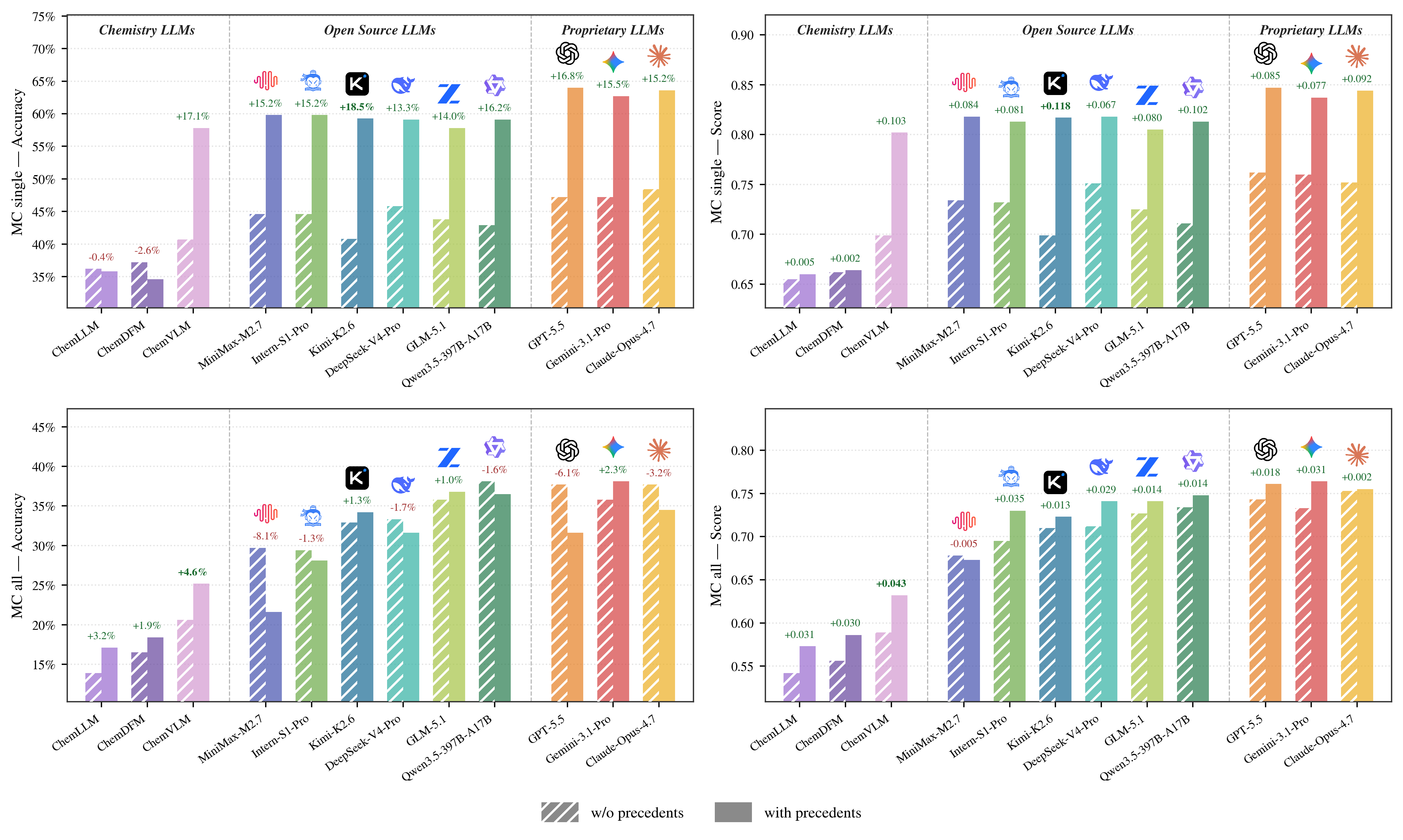}
\caption{\textbf{Precedents vs.\ no-precedents on MC.} Solid = with precedents, hatched = no-precedents; bars annotated with $\Delta_{\text{precs}}$. Top: \texttt{MC\_single}; bottom: \texttt{MC\_all}.}
\label{fig:reference_mc}
\end{figure}

\begin{figure}[t]
\centering
\includegraphics[width=\linewidth]{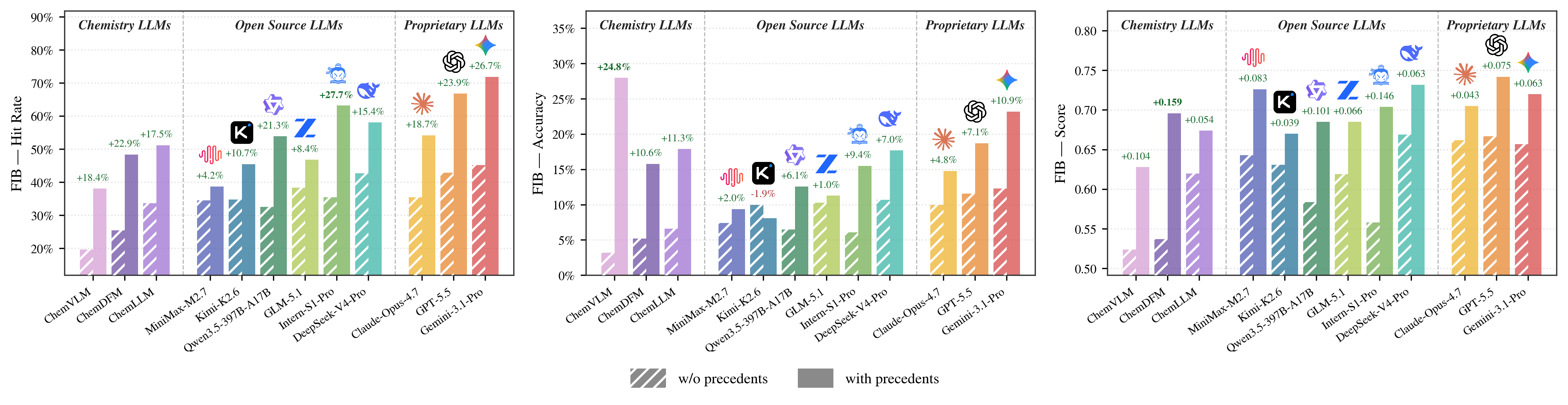}
\caption{\textbf{Precedents vs.\ no-precedents on FIB.} Solid = with precedents, hatched = no-precedents; bars annotated with $\Delta_{\text{precs}}$. Observed-row Hit, Strict Accuracy (best observed row over all submitted predictions), and Score$\,\mid\,$Hit are reported separately.}
\label{fig:reference_fib}
\end{figure}

All 12 models are evaluated on the same 4{,}773-question benchmark. Figure~\ref{fig:fullset_rose} and Table~\ref{tab:all_heatmap} report the three no-precedents subsets; Figures~\ref{fig:reference_mc}--\ref{fig:reference_fib} report the matched precedents comparisons. We highlight three top-line observations below; \S\ref{sec:perkey} and Appendix~\ref{sec:case_studies} ground them in per-key and per-reaction analyses.

\paragraph{The task is far from saturated.} No model approaches ceiling on the table-anchored selection task. The best \texttt{MC\_all} Accuracy is Claude-Opus-4.7 at $36.8\%$ (Score $0.725$); the best FIB Strict Accuracy is GPT-5.5 at $16.1\%$ (Score$\,\mid\,$Hit $0.699$); even the loosest FIB criterion---Hit, requiring the structurally valid index tuple to match an observed source-table row---peaks at $49.6\%$. For that model, the 552 outputs comprise 0 format/index-invalid responses, 278 valid but unobserved tuples, 191 observed suboptimal rows, and 83 best observed rows. Hit therefore diagnoses content-level joint-condition composition rather than output-format compliance; it does not label those non-Hits chemically infeasible. Every model also loses Score from \texttt{MC\_single} to \texttt{MC\_all} ($\Delta_{\text{axis}}\!\in\![0.030,0.090]$ for frontier LLMs), confirming that joint-condition reasoning, not single-axis ranking, is the binding constraint.

\paragraph{Chemistry-specialized models sit at the random-baseline floor.} Score must be read against the uniform-random floor of \S\ref{sec:metrics} ($\sim\!0.54$--$0.67$, top row of Table~\ref{tab:all_heatmap}), not against zero. On \texttt{MC\_all}, ChemLLM scores $20.7\%$ Accuracy / $0.545$ Score against Claude's $36.8\%$ / $0.725$---a $16.1$-pp Accuracy gap, with the Score within $0.005$ of the random floor ($0.540$). On FIB, ChemLLM reaches $6.6\%$ Strict Accuracy / $0.554$ Score$\,\mid\,$Hit against GPT-5.5's $16.1\%$ / $0.699$, only $0.014$ above the random floor ($0.540$). Thus chemistry-domain tuning yields no measurable aggregate advantage in this table-anchored decision setting. Appendix~\ref{sec:case_studies} (M2) reports the observed position-zero selection pattern as a diagnostic rather than a causal explanation.

\paragraph{Open-source LLMs close most of the gap.} The strongest open model, DeepSeek-V4-Pro, trails the strongest proprietary model by $4.0$\,pp Accuracy / $0.019$ Score on \texttt{MC\_single} ($44.6\%$ / $0.745$ vs.\ Claude $48.6\%$ / $0.764$), ties GPT-5.5 on \texttt{MC\_all} Accuracy ($32.4\%$), and reaches FIB Score$\,\mid\,$Hit $0.672$ against the top proprietary $0.683$ ($\Delta\!=\!0.011$). GLM-5.1, Qwen3.5-397B-A17B, and Kimi-K2.6 sit in the same band. The proprietary tier therefore retains a measurable but no longer decisive lead on RxnOptBench.

\subsection{Per-Condition-Key Accuracy}
\label{sec:perkey}

We break Accuracy down by the condition key each question targets, on the no-precedents subsets (MC \texttt{single\_var}, FIB \texttt{all\_blank}). The eight most frequent keys cover $933/1{,}039$ MC questions and $1{,}561/1{,}912$ FIB key occurrences; rarer keys are pooled into the headline only.

\definecolor{rowChem}{HTML}{FFE8D0}
\definecolor{rowOpen}{HTML}{DDEFE0}
\definecolor{rowClosed}{HTML}{D8E2F4}

\begin{table}[h]
\centering
\caption{Per-condition-key Accuracy (\%) on no-precedents subsets, with 12 models grouped by tier (\colorbox{rowChem}{Chemistry LLMs} / \colorbox{rowOpen}{Open Source LLMs} / \colorbox{rowClosed}{Proprietary LLMs}) and sorted ascending by overall within each tier (best at bottom). Column abbreviations: rea=reagents, sol=solvents, cat=catalysts, T=temperature, t=time, hv=light, atm=atmosphere, lig=ligand, ov=overall. \textbf{Bold} = best in column across all 12 models.}
\label{tab:perkey}
\begin{minipage}[t]{0.498\linewidth}
\centering
\textbf{(a) MC \texttt{single\_var}}\par\smallskip
\resizebox{\linewidth}{!}{%
\scriptsize
\setlength{\tabcolsep}{2.5pt}
\renewcommand{\arraystretch}{0.95}
\begin{tabular}{l rrrrrrrr | r}
\toprule
\cellcolor{blue!10}\textbf{Model} & \cellcolor{blue!14}\textbf{rea} & \cellcolor{blue!18}\textbf{sol} & \cellcolor{blue!22}\textbf{cat} & \cellcolor{blue!26}\textbf{T} & \cellcolor{blue!30}\textbf{t} & \cellcolor{blue!34}\textbf{hv} & \cellcolor{blue!38}\textbf{atm} & \cellcolor{blue!42}\textbf{lig} & \cellcolor{blue!46}\textbf{ov} \\
{\tiny $n$} & {\tiny 227} & {\tiny 201} & {\tiny 145} & {\tiny 104} & {\tiny 95} & {\tiny 73} & {\tiny 48} & {\tiny 40} & {\tiny 933} \\
\midrule
\rowcolor{rowChem}ChemLLM & 33.5 & 30.8 & 29.7 & 41.3 & 40.0 & 35.6 & 47.9 & 35.0 & 34.8 \\
\rowcolor{rowChem}ChemDFM & 34.8 & 32.8 & 29.0 & 40.4 & 42.1 & 37.0 & 50.0 & 32.5 & 35.7 \\
\rowcolor{rowChem}ChemVLM & 43.6 & 34.8 & 37.2 & 41.3 & 45.3 & 50.7 & 56.2 & 37.5 & 41.6 \\
\midrule
\rowcolor{rowOpen}Kimi-K2.6 & 40.1 & 42.3 & 34.5 & 44.2 & 42.1 & 43.8 & 43.8 & 27.5 & 40.3 \\
\rowcolor{rowOpen}Qwen3.5-397B-A17B & \textbf{45.8} & 35.8 & 33.8 & 41.3 & 44.2 & 57.5 & 45.8 & 45.0 & 42.0 \\
\rowcolor{rowOpen}Intern-S1-Pro & 44.9 & 39.8 & 30.3 & 44.2 & 35.8 & 61.6 & 56.2 & 47.5 & 42.6 \\
\rowcolor{rowOpen}MiniMax-M2.7 & 39.2 & 39.8 & 38.6 & 49.0 & 46.3 & 58.9 & 47.9 & 32.5 & 42.8 \\
\rowcolor{rowOpen}DeepSeek-V4-Pro & 42.3 & 38.8 & 35.9 & \textbf{52.9} & 47.4 & 61.6 & 50.0 & 42.5 & 44.2 \\
\rowcolor{rowOpen}GLM-5.1 & 42.3 & \textbf{42.8} & 41.4 & 49.0 & 42.1 & 56.2 & 50.0 & 42.5 & 44.5 \\
\midrule
\rowcolor{rowClosed}GPT-5.5 & 43.2 & 42.3 & 40.7 & 50.0 & 41.1 & 61.6 & 41.7 & 55.0 & 45.0 \\
\rowcolor{rowClosed}Gemini-3.1-Pro & 42.7 & 40.3 & \textbf{42.8} & 46.2 & 45.3 & \textbf{64.4} & 56.2 & 50.0 & 45.6 \\
\rowcolor{rowClosed}Claude-Opus-4.7 & 44.9 & 42.3 & \textbf{42.8} & 45.2 & \textbf{52.6} & \textbf{64.4} & \textbf{64.6} & \textbf{57.5} & \textbf{47.9} \\
\bottomrule
\end{tabular}}
\end{minipage}\hfill
\begin{minipage}[t]{0.498\linewidth}
\centering
\textbf{(b) FIB \texttt{all\_blank}}\par\smallskip
\resizebox{\linewidth}{!}{%
\scriptsize
\setlength{\tabcolsep}{2.5pt}
\renewcommand{\arraystretch}{0.95}
\begin{tabular}{l rrrrrrrr | r}
\toprule
\cellcolor{blue!10}\textbf{Model} & \cellcolor{blue!14}\textbf{rea} & \cellcolor{blue!18}\textbf{sol} & \cellcolor{blue!22}\textbf{cat} & \cellcolor{blue!26}\textbf{T} & \cellcolor{blue!30}\textbf{t} & \cellcolor{blue!34}\textbf{hv} & \cellcolor{blue!38}\textbf{atm} & \cellcolor{blue!42}\textbf{lig} & \cellcolor{blue!46}\textbf{ov} \\
{\tiny $n$} & {\tiny 334} & {\tiny 263} & {\tiny 204} & {\tiny 194} & {\tiny 202} & {\tiny 139} & {\tiny 158} & {\tiny 67} & {\tiny 1561} \\
\midrule
\rowcolor{rowChem}ChemLLM & 27.5 & 25.5 & 24.0 & 35.1 & 32.2 & 25.9 & 32.3 & 16.4 & 28.1 \\
\rowcolor{rowChem}ChemVLM & 34.7 & 38.0 & 32.4 & 47.9 & 39.1 & 40.3 & 40.5 & 44.8 & 38.7 \\
\rowcolor{rowChem}ChemDFM & 39.2 & 36.9 & 35.3 & 44.3 & 41.6 & 42.4 & 53.2 & 34.3 & 40.7 \\
\midrule
\rowcolor{rowOpen}Qwen3.5-397B-A17B & 43.4 & 45.2 & 34.8 & 49.0 & \textbf{51.5} & 46.8 & 45.6 & 43.3 & 44.8 \\
\rowcolor{rowOpen}Intern-S1-Pro & 40.7 & 44.1 & 44.1 & 48.5 & 42.6 & 51.8 & 52.5 & 52.2 & 45.6 \\
\rowcolor{rowOpen}GLM-5.1 & \textbf{47.3} & \textbf{53.2} & 41.2 & 50.5 & 49.0 & 54.7 & 47.5 & 41.8 & 48.6 \\
\rowcolor{rowOpen}MiniMax-M2.7 & 46.7 & 46.0 & 45.1 & 53.6 & 46.5 & 56.8 & 51.9 & 50.7 & 48.8 \\
\rowcolor{rowOpen}DeepSeek-V4-Pro & 47.0 & 46.4 & 44.6 & 52.6 & 40.1 & \textbf{60.4} & 57.0 & 50.7 & 48.8 \\
\rowcolor{rowOpen}Kimi-K2.6 & 46.4 & 44.5 & 48.5 & \textbf{55.7} & 46.5 & 51.1 & 56.3 & 44.8 & 48.9 \\
\midrule
\rowcolor{rowClosed}Claude-Opus-4.7 & 45.8 & 43.3 & 47.5 & 50.0 & 49.5 & 52.5 & 55.7 & 49.3 & 48.4 \\
\rowcolor{rowClosed}GPT-5.5 & 47.0 & 51.7 & \textbf{54.4} & 53.6 & 43.1 & 51.1 & 47.5 & 52.2 & 49.7 \\
\rowcolor{rowClosed}Gemini-3.1-Pro & \textbf{47.3} & 51.7 & 49.0 & 52.1 & 43.6 & 51.1 & \textbf{57.6} & \textbf{53.7} & \textbf{50.1} \\
\bottomrule
\end{tabular}}
\end{minipage}
\end{table}

\paragraph{Observations.}
\textbf{(i) The proprietary-tier advantage is concentrated on small-arity keys.} Leads exceed $20$\,pt for irradiation/light source and reaction atmosphere (typically 2--3 candidates) but compress to $\leq\!12$\,pt for reagents and reaction time, with Qwen3.5 even outperforming Claude on reagents --- the tier gap in Table~\ref{tab:all_heatmap} is largely a discrimination advantage on easy keys, not a deep-chemistry one. \textbf{(ii) Catalysts are a persistent cross-format bottleneck.} This category sits near the bottom in both MC and FIB and across all three tiers. The result establishes category-specific difficulty, but does not distinguish catalyst-class knowledge limitations from candidate ambiguity or other causes. \textbf{(iii) FIB difficulty lives in composition, not in fields.} Frontier-model overall per-key FIB Accuracy is $44.8$--$50.1\%$, while FIB Strict remains below $17\%$: individual axes carry usable signal, but models fail to compose them into the joint optimum, echoing the saturation gap reported in \S\ref{sec:results}.

\section{Conclusion and Future Work}

RxnOptBench turns reaction-condition screening into a benchmark anchored in real wet-lab tables, paired across precedents/no-precedents, and scored with a declared headline aggregate utility over reported yield and stereoselectivity. Across 12 frontier and chemistry-specialised models the task is far from saturated: chemistry-specialised models sit at the random-baseline floor, open-source LLMs close most of the proprietary lead on \texttt{MC\_single} but not on \texttt{MC\_all} or FIB, and the illustrative cases in Appendix~\ref{sec:case_studies} expose item-level patterns that complement aggregate accuracy. The results point to substantial difficulty in ordering disclosed candidates and composing multiple condition axes, rather than identifying a single universal cause of model error.

\paragraph{Future work.}
RxnOptBench can grow through versioned, expert-verified releases rather than
unbounded automatic ingestion. The automatic stages support batching,
concurrency, deterministic checks, and resumable intermediates, while manual
verification remains the deliberate quality-control bottleneck. Periodic,
versioned refreshes from newly published methodology papers can test transfer
to newly disclosed chemistry; alongside each refresh we will broaden undersampled condition families
and non-English venues and tighten FIB grading with an order-invariant chemical
comparator.

\bibliographystyle{unsrtnat}
\bibliography{references}

@phdthesis{lowe2012extraction,
  title  = {Extraction of chemical structures and reactions from the literature},
  author = {Lowe, Daniel Mark},
  year   = {2012},
  school = {University of Cambridge}
}

@article{schneider2016whats,
  title   = {What's What: The (Nearly) Definitive Guide to Reaction Role Assignment},
  author  = {Schneider, Nadine and Stiefl, Nikolaus and Landrum, Gregory A.},
  journal = {Journal of Chemical Information and Modeling},
  volume  = {56},
  number  = {12},
  pages   = {2336--2346},
  year    = {2016}
}

@article{liu2017retrosynthetic,
  title   = {Retrosynthetic reaction prediction using neural sequence-to-sequence models},
  author  = {Liu, Bowen and Ramsundar, Bharath and Kawthekar, Prasad and Shi, Jade and Gomes, Joseph and Luu Nguyen, Quang and Ho, Stephen and Sloane, Jack and Wender, Paul and Pande, Vijay},
  journal = {ACS Central Science},
  volume  = {3},
  number  = {10},
  pages   = {1103--1113},
  year    = {2017}
}

@misc{nextmove2017pistachio,
  title  = {Pistachio},
  author = {{NextMove Software}},
  year   = {2017},
  howpublished = {\url{https://www.nextmovesoftware.com/pistachio.html}}
}

@article{kearnes2021open,
  title   = {The Open Reaction Database},
  author  = {Kearnes, Steven M. and Maser, Michael R. and Wleklinski, Michael and Kast, Anton and Doyle, Abigail G. and Dreher, Spencer D. and Hawkins, Joel M. and Jensen, Klavs F. and Coley, Connor W.},
  journal = {Journal of the American Chemical Society},
  volume  = {143},
  number  = {45},
  pages   = {18820--18826},
  year    = {2021}
}

@article{wigh2024orderly,
  title   = {Orderly: data sets and benchmarks for chemical reaction data},
  author  = {Wigh, Daniel S and Arrowsmith, Joe and Pomberger, Alexander and Felton, Kobi C and Lapkin, Alexei A},
  journal = {Journal of Chemical Information and Modeling},
  volume  = {64},
  number  = {9},
  pages   = {3790--3798},
  year    = {2024},
  publisher = {ACS Publications}
}

@article{maser2021multilabel,
  title   = {Multilabel Classification Models for the Prediction of Cross-Coupling Reaction Conditions},
  author  = {Maser, Michael R. and Cui, Alexander Y. and Ryou, Serim and DeLano, Travis J. and Yue, Yisong and Reisman, Sarah E.},
  journal = {Journal of Chemical Information and Modeling},
  volume  = {61},
  number  = {1},
  pages   = {156--166},
  year    = {2021},
  publisher = {ACS Publications}
}

@inproceedings{guo2023chemllmbench,
  title     = {What can Large Language Models do in chemistry? {A} comprehensive benchmark on eight tasks},
  author    = {Guo, Taicheng and Guo, Kehan and Nan, Bozhao and Liang, Zhenwen and Guo, Zhichun and Chawla, Nitesh V. and Wiest, Olaf and Zhang, Xiangliang},
  booktitle = {Advances in Neural Information Processing Systems (NeurIPS) Datasets and Benchmarks Track},
  year      = {2023}
}

@article{mirza2025chembench,
  title     = {A framework for evaluating the chemical knowledge and reasoning abilities of large language models against the expertise of chemists},
  author    = {Mirza, Adrian and Alampara, Nawaf and Kunchapu, Sreekanth and R{\'\i}os-Garc{\'\i}a, Marti{\~n}o and Emoekabu, Benedict and Krishnan, Aswanth and Gupta, Tanya and Schilling-Wilhelmi, Mara and Okereke, Macjonathan and Aneesh, Anagha and others},
  journal   = {Nature Chemistry},
  volume    = {17},
  number    = {7},
  pages     = {1027--1034},
  year      = {2025},
  publisher = {Nature Publishing Group UK London}
}

@article{zhang2024chemllm,
  title   = {ChemLLM: A Chemical Large Language Model},
  author  = {Zhang, Di and Liu, Wei and Tan, Qian and Chen, Jingdan and Yan, Hang and Yan, Yuliang and Li, Jiatong and Huang, Weiran and Yue, Xiangyu and Zhou, Dongzhan and Zhang, Shufei and Su, Mao and Zhong, Hansen and Li, Yuqiang and Ouyang, Wanli},
  journal = {arXiv preprint arXiv:2402.06852},
  year    = {2024}
}

@inproceedings{zhao2024chemdfm,
  title     = {Chemdfm: A large language foundation model for chemistry},
  author    = {Zhao, Zihan and Ma, Da and Chen, Lu and Sun, Liangtai and Li, Zihao and Xia, Yi and Xu, Hongshen and Zhu, Zichen and Zhu, Su and Fan, Shuai and others},
  booktitle = {Neurips 2024 Workshop Foundation Models for Science: Progress, Opportunities, and Challenges},
  year      = {2024}
}

@inproceedings{li2025chemvlm,
  title     = {Chemvlm: Exploring the power of multimodal large language models in chemistry area},
  author    = {Li, Junxian and Zhang, Di and Wang, Xunzhi and Hao, Zeying and Lei, Jingdi and Tan, Qian and Zhou, Cai and Liu, Wei and Yang, Yaotian and Xiong, Xinrui and others},
  booktitle = {Proceedings of the AAAI Conference on Artificial Intelligence},
  volume    = {39},
  pages     = {415--423},
  year      = {2025}
}

@article{bran2024chemcrow,
  title   = {Augmenting large language models with chemistry tools},
  author  = {Bran, Andres M. and Cox, Sam and Schilter, Oliver and Baldassari, Carlo and White, Andrew D. and Schwaller, Philippe},
  journal = {Nature Machine Intelligence},
  volume  = {6},
  number  = {5},
  pages   = {525--535},
  year    = {2024}
}

@article{fang2024molinstructions,
  title   = {Mol-Instructions: A Large-Scale Biomolecular Instruction Dataset for Large Language Models},
  author  = {Fang, Yin and Liang, Xiaozhuan and Zhang, Ningyu and Liu, Kangwei and Huang, Rui and Chen, Zhuo and Fan, Xiaohui and Chen, Huajun},
  journal = {International Conference on Learning Representations (ICLR)},
  year    = {2024}
}

@inproceedings{wang2024scibench,
  title     = {SCIBENCH: evaluating college-level scientific problem-solving abilities of large language models},
  author    = {Wang, Xiaoxuan and Hu, Ziniu and Lu, Pan and Zhu, Yanqiao and Zhang, Jieyu and Subramaniam, Satyen and Loomba, Arjun R and Zhang, Shichang and Sun, Yizhou and Wang, Wei},
  booktitle = {Proceedings of the 41st International Conference on Machine Learning},
  pages     = {50622--50649},
  year      = {2024}
}

@article{zhou2025chemtable,
  title   = {Benchmarking Multimodal LLMs on Recognition and Understanding over Chemical Tables},
  author  = {Zhou, Yong and Chen, Mu and Mao, Qing and Luo, Yang and Liu, Qing and Li, Yi and Zhang, Xinhao and Liu, Dake and Li, Xinglong and Chen, Enhong},
  journal = {arXiv preprint arXiv:2506.11375},
  year    = {2025}
}

@misc{landrum2024rdkit,
  title  = {RDKit: Open-source cheminformatics},
  author = {Landrum, Greg},
  year   = {2024},
  howpublished = {\url{https://www.rdkit.org}}
}

@article{rajan2023decimer,
  title   = {DECIMER.ai: an open platform for automated optical chemical structure identification, segmentation and recognition in scientific publications},
  author  = {Rajan, Kohulan and Brinkhaus, Henning Otto and Agea, M. Isabel and Zielesny, Achim and Steinbeck, Christoph},
  journal = {Nature Communications},
  volume  = {14},
  pages   = {5045},
  year    = {2023}
}

@article{yang2026mineruchem,
  author  = {Yang, Haote and Wu, Jiang and Wang, Jingchao and Wei, Xingjian and Ma, Lixin and Li, Linye and Zhu, Chen and Wu, Xiaolong and Lu, Yuheng and Zhu, Ziran and others},
  title   = {{MinerU.Chem}: A High-Precision System for Optical Chemical Structure and Reaction Recognition},
  journal = {arXiv preprint arXiv:2608.03525},
  year    = {2026},
  doi     = {10.48550/arXiv.2608.03525},
  url     = {https://arxiv.org/abs/2608.03525}
}

@article{rajan2021decimerseg,
  title   = {DECIMER-Segmentation: Automated extraction of chemical structure depictions from scientific literature},
  author  = {Rajan, Kohulan and Brinkhaus, Henning Otto and Sorokina, Maria and Zielesny, Achim and Steinbeck, Christoph},
  journal = {Journal of Cheminformatics},
  volume  = {13},
  pages   = {20},
  year    = {2021}
}

@misc{anthropic2025claudeopus,
  title        = {{Claude Opus 4.7}},
  author       = {{Anthropic}},
  year         = {2026},
  howpublished = {\url{https://www.anthropic.com}}
}

@misc{openai2025gpt55,
  title        = {{GPT-5.5}},
  author       = {{OpenAI}},
  year         = {2026},
  howpublished = {\url{https://openai.com}}
}

@misc{google2025gemini31,
  title        = {{Gemini 3.1 Pro}},
  author       = {{Google DeepMind}},
  year         = {2026},
  howpublished = {\url{https://deepmind.google}}
}

@misc{deepseek2025v4,
  title        = {{DeepSeek-V4}},
  author       = {{DeepSeek-AI}},
  year         = {2026},
  howpublished = {\url{https://www.deepseek.com}}
}

@misc{moonshot2025kimik26,
  title        = {{Kimi K2.6}},
  author       = {{Moonshot AI}},
  year         = {2026},
  howpublished = {\url{https://kimi.moonshot.cn}}
}

@misc{zhipu2025glm51,
  title        = {{GLM 5.1}},
  author       = {{Zhipu AI}},
  year         = {2026},
  howpublished = {\url{https://www.zhipuai.cn}}
}

@misc{qwen2025team,
  title        = {{Qwen3} Technical Report},
  author       = {{Qwen Team}},
  year         = {2025},
  howpublished = {\url{https://qwenlm.github.io}}
}

@article{vaucher2021inferring,
  title   = {Inferring experimental procedures from text-based representations of chemical reactions},
  author  = {Vaucher, Alain C. and Schwaller, Philippe and Geluykens, Joppe and Nair, Vishnu H. and Iuliano, Anna and Laino, Teodoro},
  journal = {Nature Communications},
  volume  = {12},
  pages   = {2573},
  year    = {2021}
}

@article{wang2023parrot,
  title={Generic Interpretable Reaction Condition Predictions with Open Reaction Condition Datasets and Unsupervised Learning of Reaction Center},
  author={Wang, Xiaorui and Hsieh, Chang-Yu and Yin, Xiaodan and Wang, Jike and Li, Yuquan and Deng, Yafeng and Jiang, Dejun and Wu, Zhenxing and Du, Hongyan and Chen, Hongming and others},
  journal={Research},
  volume={6},
  pages={0231},
  year={2023},
  doi={10.34133/research.0231}
}

@article{andronov2023reagent,
  title={Reagent prediction with a molecular transformer improves reaction data quality},
  author={Andronov, Mikhail and Voinarovska, Varvara and Andronova, Natalia and Wand, Michael and Clevert, Djork-Arn{\'e} and Schmidhuber, J{\"u}rgen},
  journal={Chemical Science},
  volume={14},
  pages={3235--3246},
  year={2023},
  doi={10.1039/D2SC06798F}
}

@inproceedings{qian2023textreact,
  title={Predictive Chemistry Augmented with Text Retrieval},
  author={Qian, Yujie and Li, Zhening and Tu, Zhengkai and Coley, Connor W. and Barzilay, Regina},
  booktitle={Proceedings of the 2023 Conference on Empirical Methods in Natural Language Processing (EMNLP)},
  pages={12731--12745},
  year={2023}
}

@article{zhang2024chemmarc,
  title={Text-Augmented Multimodal LLMs for Chemical Reaction Condition Recommendation},
  author = {Zhang, Yu and Yu, Ruijie and Zeng, Kaipeng and Li, Ding and Zhu, Feng and Yang, Xiaokang and Jin, Yaohui and Xu, Yanyan},
  journal={arXiv preprint arXiv:2407.15141},
  year={2024}
}

@article{yang2025chemmas,
  title={From What to Why: A Multi-Agent System for Evidence-based Chemical Reaction Condition Reasoning},
  author={Yang, Cheng and Lu, Jiaxuan and Wan, Haiyuan and Yu, Junchi and Qin, Feiwei},
  journal={arXiv preprint arXiv:2509.23768},
  year={2025}
}

@article{garridotorres2022edboplus,
  title={A Multi-Objective Active Learning Platform and Web App for Reaction Optimization},
  author={Garrido Torres, Jose A. and Lau, Sii Hong and Anchuri, Pranay and Stevens, Jason M. and Tabora, Jose E. and Li, Jun and Borovika, Alina and Adams, Ryan P. and Doyle, Abigail G.},
  journal={Journal of the American Chemical Society},
  volume={144},
  number={43},
  pages={19999--20007},
  year={2022},
  doi={10.1021/jacs.2c08592}
}

@article{wang2024banditgeneral,
  title={Identifying general reaction conditions by bandit optimization},
  author={Wang, Jason Y. and Stevens, Jason M. and Kariofillis, Stavros K. and Tom, Mai-Jan and Golden, Dung L. and others and Doyle, Abigail G.},
  journal={Nature},
  volume={626},
  number={8001},
  pages={1025--1033},
  year={2024},
  doi={10.1038/s41586-024-07021-y}
}

@article{sin2025minerva,
  title={Highly parallel optimisation of chemical reactions through automation and machine intelligence},
  author={Sin, Joshua W. and Chau, Siu Lun and Burwood, Ryan P. and P{\"u}ntener, Kurt and Bigler, Raphael and Schwaller, Philippe},
  journal={Nature Communications},
  volume={16},
  pages={6464},
  year={2025},
  doi={10.1038/s41467-025-61803-0}
}

@inproceedings{sun2024scieval,
  title={SciEval: A Multi-Level Large Language Model Evaluation Benchmark for Scientific Research},
  author={Sun, Liangtai and Han, Yang and Zhao, Zihan and Ma, Da and Shen, Zhennan and Chen, Baocai and Chen, Lu and Yu, Kai},
  booktitle={Proceedings of the AAAI Conference on Artificial Intelligence},
  year={2024}
}

@inproceedings{lu2022scienceqa,
  title={Learn to Explain: Multimodal Reasoning via Thought Chains for Science Question Answering},
  author={Lu, Pan and Mishra, Swaroop and Xia, Tony and Qiu, Liang and Chang, Kai-Wei and Zhu, Song-Chun and Tafjord, Oyvind and Clark, Peter and Kalyan, Ashwin},
  booktitle={Advances in Neural Information Processing Systems (NeurIPS)},
  year={2022}
}

@article{yang2025mmsci,
  title   = {Does Table Source Matter? Benchmarking and Improving Multimodal Scientific Table Understanding and Reasoning},
  author  = {Yang, Bohao and Zhang, Yingji and Liu, Dong and Freitas, Andr{\'e} and Lin, Chenghua},
  journal = {arXiv preprint arXiv:2501.13042},
  year    = {2025}
}

@article{zhang2025large,
  title={Large language models to accelerate organic chemistry synthesis},
  author={Zhang, Yu and Han, Yang and Chen, Shuai and Yu, Ruijie and Zhao, Xin and Liu, Xianbin and Zeng, Kaipeng and Yu, Mengdi and Tian, Jidong and Zhu, Feng and others},
  journal={Nature Machine Intelligence},
  volume={7},
  number={7},
  pages={1010--1022},
  year={2025},
  publisher={Nature Publishing Group UK London}
}

@article{zhang2024synask,
  title={SynAsk: unleashing the power of large language models in organic synthesis},
  author={Zhang, Chonghuan and Lin, Qianghua and Zhu, Biwei and Yang, Haopeng and Lian, Xiao and Deng, Hao and Zheng, Jiajun and Liao, Kuangbiao},
  journal={Chemical science},
  volume={16},
  number={1},
  pages={43--56},
  year={2025},
  publisher={Royal Society of Chemistry}
}

@article{gao2018using,
  title={Using Machine Learning to Predict Suitable Conditions for Organic Reactions},
  author={Gao, Hanyu and Struble, Thomas J. and Coley, Connor W. and Wang, Yuran and Green, William H. and Jensen, Klavs F.},
  journal={ACS Central Science},
  volume={4},
  number={11},
  pages={1465--1476},
  year={2018}
}

@article{ahneman2018predicting,
  title={Predicting Reaction Performance in C--N Cross-Coupling Using Machine Learning},
  author={Ahneman, Derek T. and Estrada, Jes{\'u}s G. and Lin, Shishi and Dreher, Spencer D. and Doyle, Abigail G.},
  journal={Science},
  volume={360},
  number={6385},
  pages={186--190},
  year={2018}
}

@article{shields2021bayesian,
  title={Bayesian Reaction Optimization as a Tool for Chemical Synthesis},
  author={Shields, Benjamin J. and others},
  journal={Nature},
  volume={590},
  pages={89--96},
  year={2021}
}

@article{haywood2024when,
  title={When Yield Prediction Does Not Yield Prediction: An Overview of the Current Challenges},
  author={Haywood, Adam L. and others},
  journal={Journal of Chemical Information and Modeling},
  year={2024}
}

@article{boiko2023autonomous,
  title={Autonomous Chemical Research with Large Language Models},
  author={Boiko, Daniil A. and MacKnight, Robert and Kline, Ben and Gomes, Gabe},
  journal={Nature},
  volume={624},
  pages={570--578},
  year={2023}
}

@article{ramos2025review,
  title={A Review of Large Language Models and Autonomous Agents in Chemistry},
  author={Ramos, Mayk Caldas and others},
  journal={Chemical Science},
  year={2025}
}

@inproceedings{he2017maskrcnn,
  author    = {He, Kaiming and Gkioxari, Georgia and Doll{\'a}r, Piotr and Girshick, Ross},
  title     = {Mask {R-CNN}},
  booktitle = {Proceedings of the IEEE International Conference on Computer Vision (ICCV)},
  pages     = {2961--2969},
  year      = {2017},
  doi       = {10.1109/ICCV.2017.322}
}

@article{kim2025pubchem,
  author  = {Kim, Sunghwan and Chen, Jie and Cheng, Tiejun and Gindulyte, Asta and He, Jia and He, Siqian and Li, Qingliang and Shoemaker, Benjamin A. and Thiessen, Paul A. and Yu, Bo and Zaslavsky, Leonid and Zhang, Jian and Bolton, Evan E.},
  title   = {{PubChem} 2025 update},
  journal = {Nucleic Acids Research},
  volume  = {53},
  number  = {D1},
  pages   = {D1516--D1525},
  year    = {2025},
  doi     = {10.1093/nar/gkae1059}
}

@article{wang2025gtr,
  title={GTR-CoT: Graph Traversal as Visual Chain of Thought for Molecular Structure Recognition},
  author={Wang, Jingchao and He, Yifan and Yang, Haote and Wu, Jiang and Ge, Lingli and Wei, Xingjian and Wang, Yinfan and Li, Linye and Ao, Huijie and Liu, Chengjin and others},
  journal={arXiv preprint arXiv:2506.07553},
  year={2025}
}

@article{lu2022unified,
  title={Unified Deep Learning Model for Multitask Reaction Predictions with Explanation},
  author={Lu, Jieyu and Zhang, Yingkai},
  journal={Journal of Chemical Information and Modeling},
  volume={62},
  number={6},
  pages={1376--1387},
  year={2022},
  publisher={American Chemical Society},
  doi={10.1021/acs.jcim.1c01467}
}

@article{zou2026intern,
  title={Intern-S1-Pro: Scientific Multimodal Foundation Model at Trillion Scale},
  author={Zou, Yicheng and Zhu, Dongsheng and Zhu, Lin and Zhu, Tong and Zhou, Yunhua and Zhou, Peiheng and Zhou, Xinyu and Zhou, Dongzhan and Zhou, Zhiwang and Zhou, Yuhao and others},
  journal={arXiv preprint arXiv:2603.25040},
  year={2026}
}

@misc{minimax2026m27,
  author       = {{MiniMax Team}},
  title        = {MiniMax M2.7: Early Echoes of Self-Evolution},
  year         = {2026},
  month        = mar,
  howpublished = {\url{https://www.minimax.io/news/minimax-m27-en}},
  note         = {MiniMax Official Blog}
}

\newpage
\appendix

\section{Question Sub-Types and Worked Examples}
\label{app:subtypes}

See Table~\ref{tab:subtypes} of the main paper for the formal sub-type
definitions and per-sub-type counts. We work through one MC item
(\texttt{MC\_single}; dataset type \texttt{single\_varying}) and one FIB item
(\texttt{all\_blank}) below; the \texttt{MC\_all} variants share the same payload structure with the
candidate set widened to all varying axes, and the with-precedents and
matched-control variants of each sub-type share the same item structure,
differing only by the presence of a \emph{Reference Conditions And
Outcomes} block (cf.\ Appendix~\ref{app:prompts}).

\paragraph{Formal candidate sets.}
Let $\mathcal R_t$ be the distinct observed rows of table $t$ and
$\mathcal K_t$ its varying keys. For \texttt{MC\_single}, target $k$ varies
within $\{r\in\mathcal R_t:\pi_{\mathcal K_t\setminus\{k\}}(r)=c\}$, where
the observed tuple $c$ maximizes distinct target-value coverage.
\texttt{MC\_all} uses the distinct projections
$\{\pi_{\mathcal K_t}(r):r\in\mathcal R_t\}$. FIB exposes marginal lists
$\mathcal C_{q,k}=\{\pi_k(r):r\in\mathcal R_t\}$, but its observed-row
support is only $\mathcal L_q=\{\pi_{\mathcal K_t}(r):r\in\mathcal R_t\}
\subseteq\prod_k\mathcal C_{q,k}$.

\subsection{Worked example: \texttt{MC\_single}}
\label{app:sub:ex_single_mc}

We illustrate \texttt{MC\_single} with a high-discrimination item drawn
from \emph{Org.\ Lett.} \texttt{10.1021/acs.orglett.4c04736} (Wang \emph{et al.},
2025), question id
\texttt{10\_1021\_acs\_orglett\_4c04736\_1\_table\_0\_q2}. The full
question payload (input, options, and gold label) is included in the public
human-reviewed test set under this question id.

\paragraph{Reaction (constant background).}
Pd-catalysed 3,4-addition of pyrazole to a phenyl-carbamate-protected
1,2-dihydropyridine, with all conditions except the ligand held fixed:
catalyst $[\mathrm{Pd}(\eta^{3}\text{-}\mathrm{C}_{3}\mathrm{H}_{5})\mathrm{Cl}]_{2}$,
solvent PhCF$_{3}$, temperature 50\,$^{\circ}$C.

\paragraph{Target axis and options.}
The model is asked to choose the optimal value of the single varying axis
\texttt{ligand} from four candidates, all four of which are real entries in
Table~1 of the cited paper:
\texttt{[0] BINAP}, \texttt{[1] DPPF}, \texttt{[2] Dpephos}, \texttt{[3] PPh$_3$}.
Their measured target-product yields are 35\%, 21\%, 96\%, and 0\%,
respectively. \textbf{Gold:} \texttt{[2] Dpephos}; the next-best option,
BINAP, has relative score $r\!\approx\!0.36$.

\paragraph{Model behaviour.}
The three Proprietary LLMs, \texttt{ChemVLM}, and
\texttt{DeepSeek-V4-Pro} recover the gold ligand. \texttt{ChemLLM},
\texttt{ChemDFM}, \texttt{GLM-5.1}, and \texttt{Kimi-K2.6} choose BINAP;
\texttt{MiniMax-M2.7} chooses DPPF; and \texttt{Intern-S1-Pro} and
\texttt{Qwen3.5-397B-A17B} choose PPh$_3$. The item therefore
discriminates among closely related ligand choices whose target-product
yields differ substantially. Aggregated per-model behaviour across all
sub-types is reported in Section~\ref{sec:results}.

\subsection{Worked example: FIB \texttt{all\_blank}}
\label{app:sub:ex_fib}

We illustrate FIB \texttt{all\_blank} with a 3-key item drawn from
\emph{Org.\ Lett.} \texttt{10.1021/acs.orglett.5c02358} (SI Table S2, p.\ S14),
question id
\texttt{10\_1021\_acs\_orglett\_5c02358\_si\_13\_table\_0\_q1}. The full
question payload (input, candidate values, and gold label) is included in the
public human-reviewed test set under this question id.

\paragraph{Reaction (constant background).}
Photoredox spirocyclopropanation of \emph{N}-phenyl oxamic acid (\textbf{1a})
with methyl 2-(chloromethyl)acrylate (\textbf{2a}), with photocatalyst
4CzIPN (8\,mol\%), base KOH, light source 455\,nm LED, atmosphere N$_2$.

\paragraph{Query keys and candidate values.}
The model is asked to fill three blanks simultaneously, each constrained to
a small enumerated candidate list: temperature
$\{35,40\}\,^{\circ}$C, time $\{24,12\}$\,h, and solvent
$\{\mathrm{DMF},\mathrm{DMSO}\}$. The Cartesian product contains eight
tuples, while retained measured-row support contains four:
\texttt{(35\,$^{\circ}$C,\ 24\,h,\ DMF)} at 26\%,
\texttt{(35\,$^{\circ}$C,\ 24\,h,\ DMSO)} at 80\%,
\texttt{(35\,$^{\circ}$C,\ 12\,h,\ DMSO)} at 36\%, and
\texttt{(40\,$^{\circ}$C,\ 24\,h,\ DMSO)} at 57\%.

\textbf{Gold:} \texttt{(35\,$^{\circ}$C,\ 24\,h,\ DMSO)} with 80\%
yield. The other four Cartesian combinations are well-formed but do not
correspond to a retained measured row, so they are non-Hits and receive no
Score$\,\mid\,$Hit. This example directly illustrates why marginally valid
field choices need not compose into an observed tuple.

\paragraph{Model behaviour.}
All three Proprietary LLMs plus \texttt{kimi-k2.6} recover the gold
combination. \texttt{GLM-5.1}, \texttt{MiniMax-M2.7}, and
\texttt{Qwen3.5-397B-A17B} select the observed but suboptimal
\texttt{(35\,$^{\circ}$C,\ 24\,h,\ DMF)} tuple. \texttt{ChemDFM},
\texttt{ChemVLM}, and \texttt{DeepSeek-V4-Pro} select the unobserved
\texttt{(40\,$^{\circ}$C,\ 24\,h,\ DMF)} tuple, while
\texttt{ChemLLM} and \texttt{Intern-S1-Pro} select the unobserved
\texttt{(40\,$^{\circ}$C,\ 12\,h,\ DMSO)} tuple. Thus the errors separate
suboptimal observed-row selection from recombinations outside observed
support.

\section{Human Verification and Anti-Leakage Safeguards}
\label{app:safeguards}

\subsection{Annotation interface}
\label{app:safe:ui}

The per-table review interface operationalises the expert-review gate of the
extraction pipeline (Section~\ref{sec:pipeline}). During local review, a
lawfully accessed source crop is displayed in a non-released pane beside
editable structured fields for reactants and products (content names, SMILES,
\texttt{idt} labels, amounts, yields, conversion rate,
diastereomeric/enantiomeric ratio, and free-text \texttt{other\_info}).
The toolbar (\texttt{Accept} / \texttt{Save edits} / \texttt{Skip} /
\texttt{Reject} / \texttt{Reset} / \texttt{Prev} / \texttt{Next} /
\texttt{Next pending} / \texttt{List}) lets reviewers triage tables in
order, patch any field inline before saving, and persist the verdict
together with a timestamp. Only entries with status \emph{accepted}
flow downstream to MC and FIB question construction; \emph{rejected}
and \emph{skipped} tables are excluded from the released benchmark.

Verification was performed by six paid interns based in mainland China:
four master's-level interns received RMB~300 per working day and two
doctoral-level interns received RMB~400 per working day.

\subsection{Anti-leakage safeguards}
\label{app:safe:leakage}

Sources are restricted to papers first published in 2025, and the released
metadata retain DOI and first-publication provenance. This recency lowers but
cannot eliminate exact-source exposure: hosted and open models have
heterogeneous and partly undisclosed training cutoffs, so we do not claim a
contamination-free evaluation. All models receive the same frozen question
files and candidate permutations, and the release manifest hashes the
distributed files. Future versioned refreshes can support temporal
contamination audits.

\section{Benchmark Composition Across Reaction, Condition, and Product Dimensions}
\label{app:coverage}

\paragraph{Counting unit and audit protocol.}
The 4{,}773 benchmark questions are derived from 552 source screening tables,
so question-level counts would repeatedly count the same chemistry. We
therefore use one parent table as the unit throughout this section
($N=552$). Each table was assigned one of 20 mutually exclusive primary
transformations and independent multi-label energy-input and catalyst-class
annotations. Two blind Gemini-3.1-Pro-Low passes were followed by field-level
blinded adjudication; all 10 flagged records were manually checked against the
reviewed table record and, when needed, the publisher source. The released
audit contains all per-table decisions and leaves no unresolved record.

\begin{figure*}[t]
  \centering
  \includegraphics[width=\textwidth]{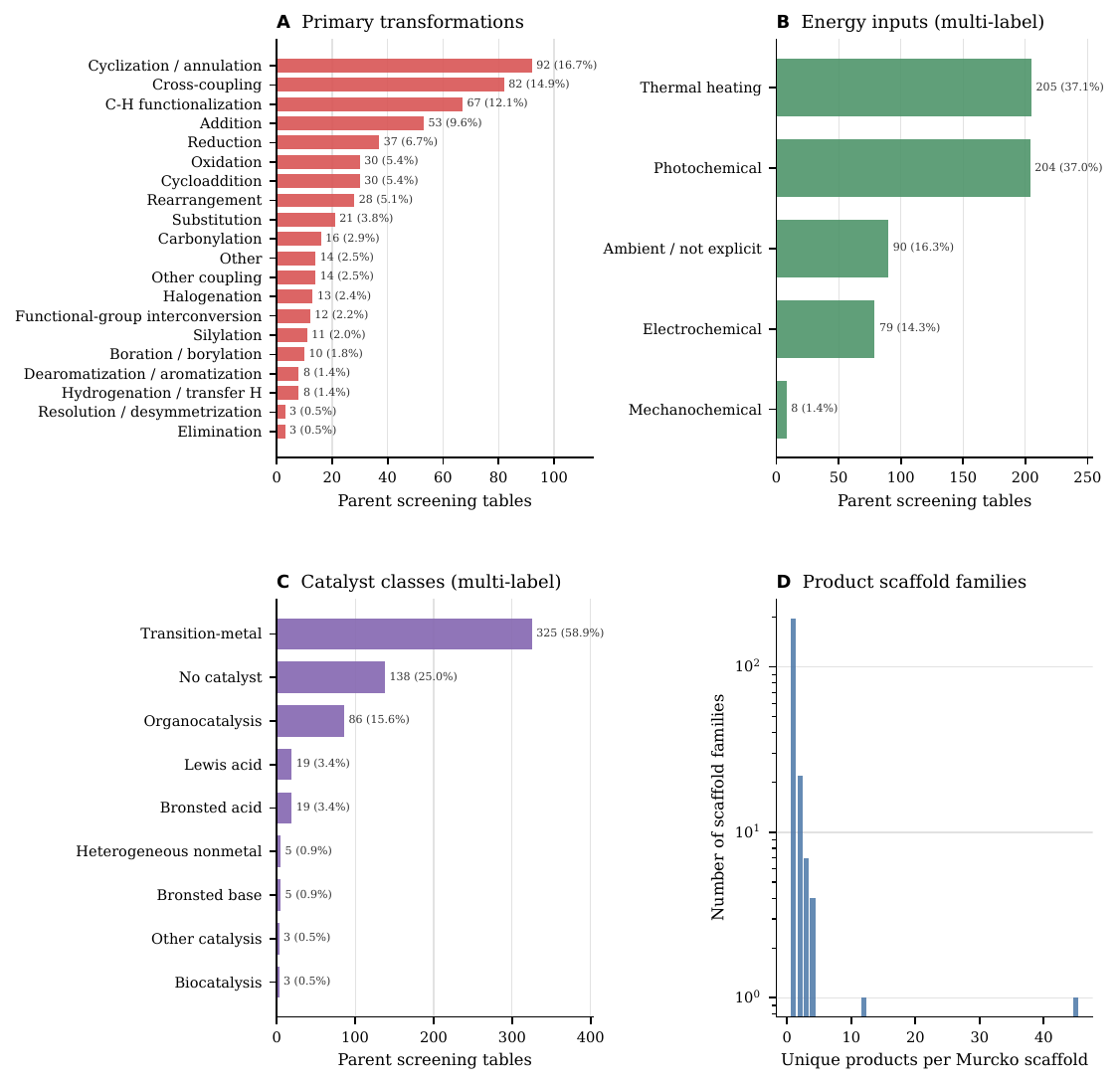}
  \caption{\textbf{Benchmark composition across reaction and product
  dimensions}. \textbf{A}, mutually exclusive primary transformation classes
  over 552 parent screening tables. \textbf{B--C}, multi-label energy inputs
  and catalyst classes; labels report table count and share of all parent
  tables. \textbf{D}, family sizes over 232 non-empty Bemis--Murcko
  scaffolds from 341 unique canonical target products.}
  \label{fig:benchmark_composition}
\end{figure*}

\paragraph{Reaction and product composition.}
Figure~\ref{fig:benchmark_composition} summarizes complementary aspects of
the benchmark chemistry. All 20 primary transformation classes are
represented, and the largest accounts for 16.7\% of parent tables. The corpus
also spans thermal, photochemical, and electrochemical activation together
with multiple catalyst regimes. All 552 tables contain a valid target-product
structure, yielding 341 unique canonical products and 232 non-empty
Bemis--Murcko scaffolds after deduplication. Radius-2 Morgan-fingerprint
internal diversity is 0.875.

\begin{figure*}[t]
  \centering
  \includegraphics[width=\textwidth]{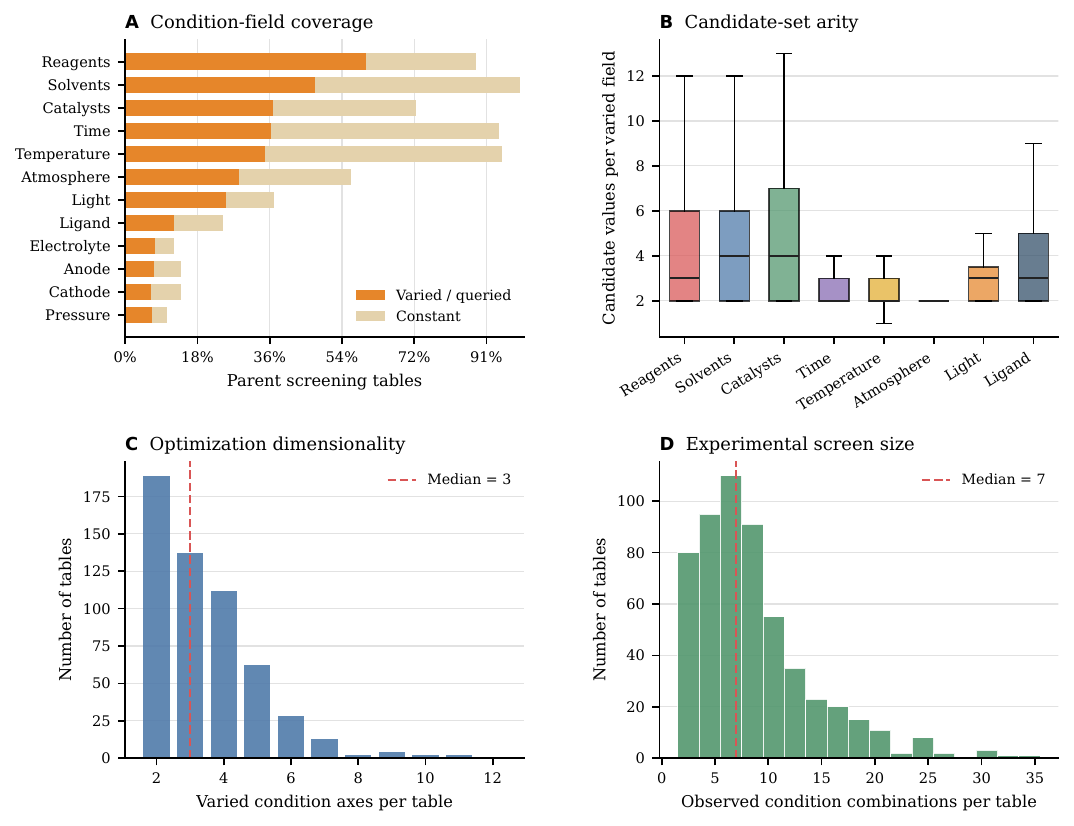}
  \caption{\textbf{Condition-space structure} ($N=552$). \textbf{A}, fraction
  of parent tables in which each major condition family is varied/queried or
  held constant. \textbf{B}, candidate-set arity for varied fields (boxes
  show quartiles; whiskers show the non-outlier range). \textbf{C}, number of
  varied condition axes per parent table. \textbf{D}, number of
  experimentally observed condition combinations per source table.}
  \label{fig:coverage_conditions}
\end{figure*}

\paragraph{Condition-space structure.}
The structured schema contains 26 condition fields
(Figure~\ref{fig:coverage_conditions}). Solvent, temperature, time, reagent,
and catalyst are the most frequently represented families. A parent table
varies a median of three condition axes and reports a median of seven
experimentally observed condition combinations, while candidate-set size
varies by field. The four panels distinguish field coverage, per-field
branching, table dimensionality, and observed screen size.

\paragraph{Scope.}
The 2025 sampling frame is a deliberate target distribution rather than an
estimate of historical publication prevalence. Recent methodology papers
feature contemporary transformations, catalytic systems, and condition
regimes, and their screening tables provide expert-designed local decision
spaces grounded in measured experiments. RxnOptBench therefore probes whether
models transfer chemical knowledge to recent, potentially unfamiliar
condition-selection problems rather than only recall canonical recipes.
Recency lowers exact source-exposure risk relative to older corpora but cannot
guarantee contamination-free evaluation, and we do not assume that every
transformation constitutes a new reaction class. Together, the analyses above
characterize substantial reaction, condition, and product heterogeneity within
this intended distribution across 552 parent screening tables.
\FloatBarrier

\section{Benchmark Validity and Robustness Audits}
\label{app:robustness}

\paragraph{Disclosed outcome variation and narrative selection.}
RxnOptBench evaluates only the candidates disclosed in each source table; it
does not interpret the table as a complete optimization history or its best
row as a global chemical optimum. Publication curation defines the sampling
frame, but it does not alter the measured outcomes or within-table ordering
used as labels. For a static offline benchmark without additional wet-lab
validation, only observed joint rows have defensible outcome scores;
unreported combinations remain outside measured support rather than becoming
failures or negative examples. On the fixed FIB task set, after format/index
invalidity is separated, lower Hit therefore diagnoses difficulty coordinating
multiple field choices into an experimentally observed joint row; it does not
adjudicate the wet-lab feasibility of non-Hits. Within this conditional task,
the disclosed outcomes are not narrowly concentrated: across the 552 parent tables, the
best-to-worst yield difference has median 69 percentage points (IQR
40.75--83; range 1--100). In total, 527/552 tables (95.5\%) span at least
10 points, 493 (89.3\%) span at least 20, and 383 (69.4\%) span at least 50.
When tables are stratified into quartiles by yield spread, the ordering of the
12 models remains positively associated with the full-set ordering
(Spearman $\rho=0.755,0.895,0.902,0.965$ from the lowest to highest
quartile), and the Proprietary, Open Source, and Chemistry model tiers retain
the same order in every quartile. Thus, publication curation bounds claims
about exhaustive coverage, while the reported model-tier conclusion is not
confined to tables with negligible disclosed outcome differences.

\paragraph{Scoring sensitivity and multi-objective view.}
The multiplicative score is the benchmark's headline aggregate utility: a
monotone, bounded summary that provides one continuous ranking while preserving
partial credit. It is not asserted to encode every chemical objective or to be
the unique preference function for experimental optimization. We therefore
re-score \texttt{MC\_all} in three ways. First, replacing the product by
yield-only scoring gives a model-rank Spearman correlation of $0.965$ across
all 12 models. Second, 43 tables report at least one selectivity axis for
every option, permitting complete objective vectors consisting of normalized
yield and all complete major-fraction selectivity axes. On these tables,
equal-weight additive and maximin utilities give rank correlations of $0.965$
and $0.923$, respectively, with the product score. Third, for those 43 tables,
we compute a per-prediction Pareto flag across all 12 models without
collapsing the objective vector. We
retain the product score for a single headline ranking and provide these
objective vectors and alternatives as complementary diagnostics. These are
post hoc sensitivity and multi-objective diagnostics for the present
single-selection task, not a claim that RxnOptBench implements full
multi-objective optimization. A full formulation would require substantially
more tables with complete candidate-level yield and selectivity reporting and
a different task definition, such as Pareto-front recovery or recommendation
under explicit preferences and constraints.

The archived evaluation already used the signed-\emph{ee} and ratio-direction
normalization defined in Section~\ref{sec:metrics}, so Table~\ref{tab:all_heatmap}
retains the archived snapshot values. Separately, the evaluation parser was
hardened to reject numeric compound labels such as \texttt{3a:4a} as ratios.
A disclosed maintenance recomputation changes only ChemLLM FIB Strict Accuracy
from the frozen 6.6\% to 5.3\%; we report this audit rather than silently
replacing the archived result.

\paragraph{Candidate-position audit.}
All models receive the same shuffled candidate order. MC bundles are
deterministically shuffled per question; FIB candidate values are
independently shuffled per question and condition key with a stable hash and
seed \texttt{20260504}. The exact permutations are stored in the public
test-set records. The observed frequency with which a gold answer includes
index 0 is 375/1{,}039 (36.09\%) for \texttt{MC\_single}, 114/552 (20.65\%)
for \texttt{MC\_all}, and 708/1{,}912 (37.03\%) across FIB key occurrences.
Because option arity and ties vary by item, their corresponding
arity-conditional uniform expectations are 36.50\%, 17.75\%, and 36.41\%;
two-sided Monte Carlo tests give $p=0.786$, $0.063$, and $0.569$,
respectively. Thus none differs significantly from the conditional shuffled
null at $0.05$. The public evaluation code reproduces the complete index
counts and percentages stratified by task and candidate arity.

\paragraph{Inference reproducibility boundary.}
Formal runs freeze datasets, prompts, candidate permutations, model IDs, and
generation settings and disable sampling (\texttt{temperature=0} or provider
greedy/non-sampling mode). The public evaluator deterministically scores any
supplied predictions against the fixed test set. Extraction code,
benchmark-construction code, and the archived model outputs used for the paper
are not part of the public release. Hosted backends may not reproduce
generations bit-for-bit; repeated complete API calls would measure
endpoint-level run-to-run variation, and we do not claim that hosted-model
responses are invariant across such calls.
\FloatBarrier

\section{License, Hosting, Intended Use, and Broader Impact}
\label{app:license}

The 232 papers and their supporting information were manually downloaded by
the authors through open-access channels or university institutional
subscriptions; no automated crawling or bulk downloading from publisher
websites was used. Processing was performed locally. Before final public
release, we will document the acquisition route and verify the applicable
publisher TDM term, permission, or legal exception for each source category;
records without a confirmed release basis will be excluded.

The release is limited to the evaluation code and final human-reviewed
benchmark test-set records, which retain DOI/table/page provenance. It does
not redistribute the extraction pipeline, benchmark-construction code, review
tooling, archived model outputs, raw PDFs, supporting-information files,
rendered page/table images, publisher-formatted text, or verbatim source text.
The test-set records will be available through a lightweight research-use
agreement requiring source attribution, compliance with applicable laws and
source terms, and acknowledgement of the intended evaluation/research use.
Final license choices, persistent identifiers, and the integrity-manifest hash
will be fixed at camera-ready time.

\paragraph{Broader impact and misuse mitigation.}
Machine-readable aggregation lowers the effort required to locate and reuse
published reaction-condition information. The records could therefore be
misused as unvalidated condition recommendations outside their source
context. Users will be instructed not to treat benchmark or model outputs as
direct wet-lab guidance without qualified expert review and independent
experimental validation. The research-use agreement, retained source
provenance, explicit intended-use statement, and a channel for reporting
misuse or rights concerns provide accountability and friction, although they
cannot eliminate misuse. Using benchmark items as training data would also
invalidate future evaluation on RxnOptBench.

\section{Prompt Templates}
\label{app:prompts}

Although RxnOptBench has eight evaluated subsets (Table~\ref{tab:subtypes}), they share \textbf{three} prompt templates: \texttt{mc\_single}, \texttt{mc\_all}, and \texttt{fib\_all}, the renderer aliases for \texttt{MC\_single}, \texttt{MC\_all}, and FIB \texttt{all\_blank}, respectively. The reference vs.\ control distinction is purely a toggle on a single \emph{Reference Conditions And Outcomes} section --- when present, 2--5 prior wet-lab rows are appended; when absent, every other section is identical. This section therefore reproduces the three full prompts in their reference variant; removing the \emph{Reference Conditions And Outcomes} block recovers the matched control prompt verbatim.

Every submitted prompt contains a one-line instruction, \emph{Reaction}
(reactants and target product as name + SMILES), and \emph{Constant
Conditions}, i.e.\ table-level invariants shared by all source-table rows.
The \texttt{mc\_single} templates additionally include \emph{Fixed
Conditions}: question-level locked conditions that may vary elsewhere in the
table but are held at one observed combination for the selected subgroup.
Reference variants add \emph{Reference Conditions And Outcomes}; each
template then gives its candidate set and output format. An audit of the actual
task templates and all eight evaluated subsets confirms that all 34,025 stored
constant-condition keys and values are rendered across the 4,773 prompts.
From the reviewer-archive root, the audit is reproduced by
\begin{verbatim}
python scripts/validate_prompt_context.py
\end{verbatim}

\subsection{\texttt{mc\_single}}
\label{app:prompt:mc_single}

Sample id: \texttt{10\_1002\_anie\_202416132\_si\_5\_table\_0\_q1\_ref}; expected output: \texttt{1}.

\begin{promptbox}{\texttt{mc\_single + reference}}
You are an expert organic chemist. Given the following reaction, one real fixed condition group from the source table, and several reference condition sets with observed outcomes, choose the best value for **Constant_Current** from the remaining candidate options.

### Reaction

**Reactants:**
  - 2,2-difluoro-3-phenyl-2H-indole (SMILES: FC1(F)C(c2ccccc2)=Nc2ccccc21)

**Target Product:**
  - 2-fluoro-3-phenyl-1H-indole (SMILES: Fc1c(-c2ccccc2)[nH]c2ccccc12)

### Constant Conditions (do not change)

  - Anode(+): Zn
  - Cathode(-): C
  - Main_Mode: Constant
  - Reaction_Type: Integrated Cell Reaction
  - reaction_temperature: rt
  - speed: 800 rpm

### Fixed Conditions

  - reaction_time: 4 h
  - reagents: PS-750-M
  - solvents: water (SMILES: O)

### Reference Conditions And Outcomes

*Reference 1 (yield=59.0%):*
  - Anode(+): Zn
  - Cathode(-): C
  - Constant_Current: 10 mA
  - Main_Mode: Constant
  - Reaction_Type: Integrated Cell Reaction
  - reaction_temperature: rt
  - reaction_time: 18 h
  - reagents: (none)
  - solvents: 3 wt % aq. PS-750-M
  - speed: 800 rpm

*Reference 2 (yield=92.0%):*
  - Anode(+): Zn
  - Cathode(-): C
  - Constant_Current: 20 mA
  - Main_Mode: Constant
  - Reaction_Type: Integrated Cell Reaction
  - reaction_temperature: rt
  - reaction_time: 4 h
  - reagents: PS-750-M
  - solvents: (none)
  - speed: 800 rpm

### Candidate Values For Constant_Current

**Option 0:**
  - Constant_Current: 10 mA

**Option 1:**
  - Constant_Current: 30 mA

### Output Format

Respond with ONLY the option index (0-based integer). Do not include any reasoning, explanation, or additional text --- just the integer.
\end{promptbox}

\subsection{\texttt{mc\_all}}
\label{app:prompt:mc_all}

Sample id: \texttt{10\_1002\_anie\_202416132\_si\_4\_table\_1\_q6\_ref}; expected output: \texttt{3}.

\begin{promptbox}{\texttt{mc\_all + reference}}
You are an expert organic chemist. Given the following reaction and several reference condition sets with observed outcomes, choose the best remaining reaction-condition option.

### Reaction

**Reactants:**
  - 2,2-difluoro-3-phenyl-2H-indole (SMILES: FC1(F)C(c2ccccc2)=Nc2ccccc21)

**Target Product:**
  - 2-fluoro-3-phenyl-1H-indole (SMILES: Fc1c(-c2ccccc2)[nH]c2ccccc12)

### Reference Conditions And Outcomes

*Reference 1 (yield=27.0%):*
  - Anode(+): Zn
  - Cathode(-): C
  - Constant_Current: 30 mA
  - Electrolyte: n-Bu4NClO4 (0.1 M)
  - Main_Mode: Constant
  - Reaction_Type: Integrated Cell Reaction
  - other_info: undivided cell, Electrolysis conducted with DC power supply using Electrasyn 2.0
  - reaction_temperature: rt
  - reaction_time: 4 h
  - reagents: n-Bu4NClO4 (SMILES: CCCC[N+](CCCC)(CCCC)CCCC.[O-][Cl+3]([O-])([O-])[O-])
  - scale: (none)
  - selected_conditions: ["reagents", "solvents", "reaction_time", "reaction_temperature", "speed", "current_conditions"]
  - solvents: 3 wt % aq. PS-750-M
  - speed: 800 rpm

*Reference 2 (yield=32.0%):*
  - Anode(+): Zn
  - Cathode(-): C
  - Constant_Current: 30 mA
  - Electrolyte: NaI (0.1 M)
  - Main_Mode: Constant
  - Reaction_Type: Integrated Cell Reaction
  - other_info: undivided cell, Electrolysis conducted with DC power supply using Electrasyn 2.0
  - reaction_temperature: rt
  - reaction_time: 4 h
  - reagents: NaI (SMILES: [I-].[Na+])
  - scale: 0.25 mmol
  - selected_conditions: ["reagents", "solvents", "reaction_time", "reaction_temperature", "speed", "current_conditions", "scale"]
  - solvents: PS-750-M
  - speed: 800 rpm

*Reference 3 (yield=42.0%):*
  - Anode(+): Zn
  - Cathode(-): C
  - Constant_Current: 30 mA
  - Electrolyte: n-Bu4NBr (0.1 M)
  - Main_Mode: Constant
  - Reaction_Type: Integrated Cell Reaction
  - other_info: undivided cell, Electrolysis conducted with DC power supply using Electrasyn 2.0
  - reaction_temperature: rt
  - reaction_time: 4 h
  - reagents: n-Bu4NBr (SMILES: CCCC[N+](CCCC)(CCCC)CCCC.[Br-])
  - scale: (none)
  - selected_conditions: ["reagents", "solvents", "reaction_time", "reaction_temperature", "speed", "current_conditions", "indeterminate_descriptions"]
  - solvents: 3 wt % aq. PS-750-M
  - speed: 800 rpm

### Candidate Conditions

**Option 0:**
  - Electrolyte: NaBr (0.1 M)
  - reagents: NaBr (SMILES: [Br-].[Na+])
  - scale: 0.25 mmol
  - selected_conditions: ["reagents", "solvents", "current_conditions", "reaction_time", "reaction_temperature", "speed", "scale"]
  - solvents: aq. PS-750-M

**Option 1:**
  - Electrolyte: n-Bu4NPF6 (0.1 M)
  - reagents: n-Bu4NPF6 (SMILES: CCCC[N+](CCCC)(CCCC)CCCC.F[P-](F)(F)(F)(F)F)
  - scale: 0.25 mmol
  - selected_conditions: ["reagents", "solvents", "reaction_time", "reaction_temperature", "speed", "current_conditions", "scale"]
  - solvents: PS-750-M

**Option 2:**
  - Electrolyte: NaBF4 (0.1 M)
  - reagents: NaBF4 (SMILES: F[B-](F)(F)F.[Na+])
  - scale: 0.25 mmol
  - selected_conditions: ["reagents", "solvents", "reaction_time", "reaction_temperature", "speed", "current_conditions", "scale"]
  - solvents: PS-750-M

**Option 3:**
  - Electrolyte: n-Bu4NI (0.1 M)
  - reagents: n-Bu4NI (SMILES: CCCC[N+](CCCC)(CCCC)CCCC.[I-])
  - scale: 0.25 mmol
  - selected_conditions: ["reagents", "solvents", "reaction_time", "reaction_temperature", "speed", "current_conditions", "scale"]
  - solvents: 3 wt % aq. PS-750-M

**Option 4:**
  - Electrolyte: NaClO4 (0.1 M)
  - reagents: NaClO4 (SMILES: [Na+].[O-][Cl+3]([O-])([O-])[O-])
  - scale: 0.25 mmol
  - selected_conditions: ["reagents", "solvents", "reaction_time", "reaction_temperature", "speed", "current_conditions", "scale"]
  - solvents: PS-750-M

### Output Format

Respond with ONLY the option index (0-based integer).
\end{promptbox}

\subsection{\texttt{fib\_all}}
\label{app:prompt:fib_all}

Sample id: \texttt{10\_1002\_anie\_202416132\_si\_4\_table\_1\_q1\_ref}; expected output: a JSON object with one integer per condition key (e.g.\ \texttt{\{"Electrolyte":\ 0,\ "reagents":\ 0,\ "scale":\ 0,\ "selected\_conditions":\ 0,\ "solvents":\ 0\}}).

\begin{promptbox}{\texttt{fib\_all + reference}}
You are an expert organic chemist. Given the following reaction, several reference condition sets with observed outcomes, and a candidate index table for each condition category, choose one candidate index for each required condition category.

### Reaction

**Reactants:**
  - 2,2-difluoro-3-phenyl-2H-indole (SMILES: FC1(F)C(c2ccccc2)=Nc2ccccc21)

**Target Product:**
  - 2-fluoro-3-phenyl-1H-indole (SMILES: Fc1c(-c2ccccc2)[nH]c2ccccc12)

### Constant Conditions

  - Anode(+): Zn
  - Cathode(-): C
  - Constant_Current: 30 mA
  - Main_Mode: Constant
  - Reaction_Type: Integrated Cell Reaction
  - other_info: undivided cell, Electrolysis conducted with DC power supply using Electrasyn 2.0
  - reaction_temperature: rt
  - reaction_time: 4 h
  - speed: 800 rpm

### Reference Conditions And Outcomes

*Reference 1 (yield=27.0%):*
  - Anode(+): Zn
  - Cathode(-): C
  - Constant_Current: 30 mA
  - Electrolyte: n-Bu4NClO4 (0.1 M)
  - Main_Mode: Constant
  - Reaction_Type: Integrated Cell Reaction
  - other_info: undivided cell, Electrolysis conducted with DC power supply using Electrasyn 2.0
  - reaction_temperature: rt
  - reaction_time: 4 h
  - reagents: n-Bu4NClO4 (SMILES: CCCC[N+](CCCC)(CCCC)CCCC.[O-][Cl+3]([O-])([O-])[O-])
  - scale: (none)
  - selected_conditions: ["reagents", "solvents", "reaction_time", "reaction_temperature", "speed", "current_conditions"]
  - solvents: 3 wt % aq. PS-750-M
  - speed: 800 rpm

*Reference 2 (yield=32.0%):*
  - Anode(+): Zn
  - Cathode(-): C
  - Constant_Current: 30 mA
  - Electrolyte: NaI (0.1 M)
  - Main_Mode: Constant
  - Reaction_Type: Integrated Cell Reaction
  - other_info: undivided cell, Electrolysis conducted with DC power supply using Electrasyn 2.0
  - reaction_temperature: rt
  - reaction_time: 4 h
  - reagents: NaI (SMILES: [I-].[Na+])
  - scale: 0.25 mmol
  - selected_conditions: ["reagents", "solvents", "reaction_time", "reaction_temperature", "speed", "current_conditions", "scale"]
  - solvents: PS-750-M
  - speed: 800 rpm

*Reference 3 (yield=42.0%):*
  - Anode(+): Zn
  - Cathode(-): C
  - Constant_Current: 30 mA
  - Electrolyte: n-Bu4NBr (0.1 M)
  - Main_Mode: Constant
  - Reaction_Type: Integrated Cell Reaction
  - other_info: undivided cell, Electrolysis conducted with DC power supply using Electrasyn 2.0
  - reaction_temperature: rt
  - reaction_time: 4 h
  - reagents: n-Bu4NBr (SMILES: CCCC[N+](CCCC)(CCCC)CCCC.[Br-])
  - scale: (none)
  - selected_conditions: ["reagents", "solvents", "reaction_time", "reaction_temperature", "speed", "current_conditions", "indeterminate_descriptions"]
  - solvents: 3 wt % aq. PS-750-M
  - speed: 800 rpm

Important: the reference condition sets shown above are NOT valid answers for this question. Do NOT copy any reference combination directly. You must propose a NEW combination by selecting candidate indices from the table below. Only combinations formed from the candidate index table are considered valid outputs.

### Candidate Values

**Electrolyte:**
  0: n-Bu4NI (0.1 M)
  1: n-Bu4NPF6 (0.1 M)
  2: NaClO4 (0.1 M)
  3: NaBr (0.1 M)
  4: NaBF4 (0.1 M)

**reagents:**
  0: n-Bu4NI (SMILES: CCCC[N+](CCCC)(CCCC)CCCC.[I-])
  1: n-Bu4NPF6 (SMILES: CCCC[N+](CCCC)(CCCC)CCCC.F[P-](F)(F)(F)(F)F)
  2: NaClO4 (SMILES: [Na+].[O-][Cl+3]([O-])([O-])[O-])
  3: NaBr (SMILES: [Br-].[Na+])
  4: NaBF4 (SMILES: F[B-](F)(F)F.[Na+])

**scale:**
  0: 0.25 mmol

**selected_conditions:**
  0: ["reagents", "solvents", "reaction_time", "reaction_temperature", "speed", "current_conditions", "scale"]
  1: ["reagents", "solvents", "current_conditions", "reaction_time", "reaction_temperature", "speed", "scale"]

**solvents:**
  0: 3 wt % aq. PS-750-M
  1: PS-750-M
  2: aq. PS-750-M

### Output Format

Respond strictly in the following JSON format (no extra text):

```json
{
  "Electrolyte": <candidate_index>,
  "reagents": <candidate_index>,
  "scale": <candidate_index>,
  "selected_conditions": <candidate_index>,
  "solvents": <candidate_index>
}
```

Each value must be a single integer candidate index, not a condition name.
\end{promptbox}

\section{Illustrative Failure Cases}
\label{sec:case_studies}

\begin{figure}[ht]
\centering
\includegraphics[width=\linewidth]{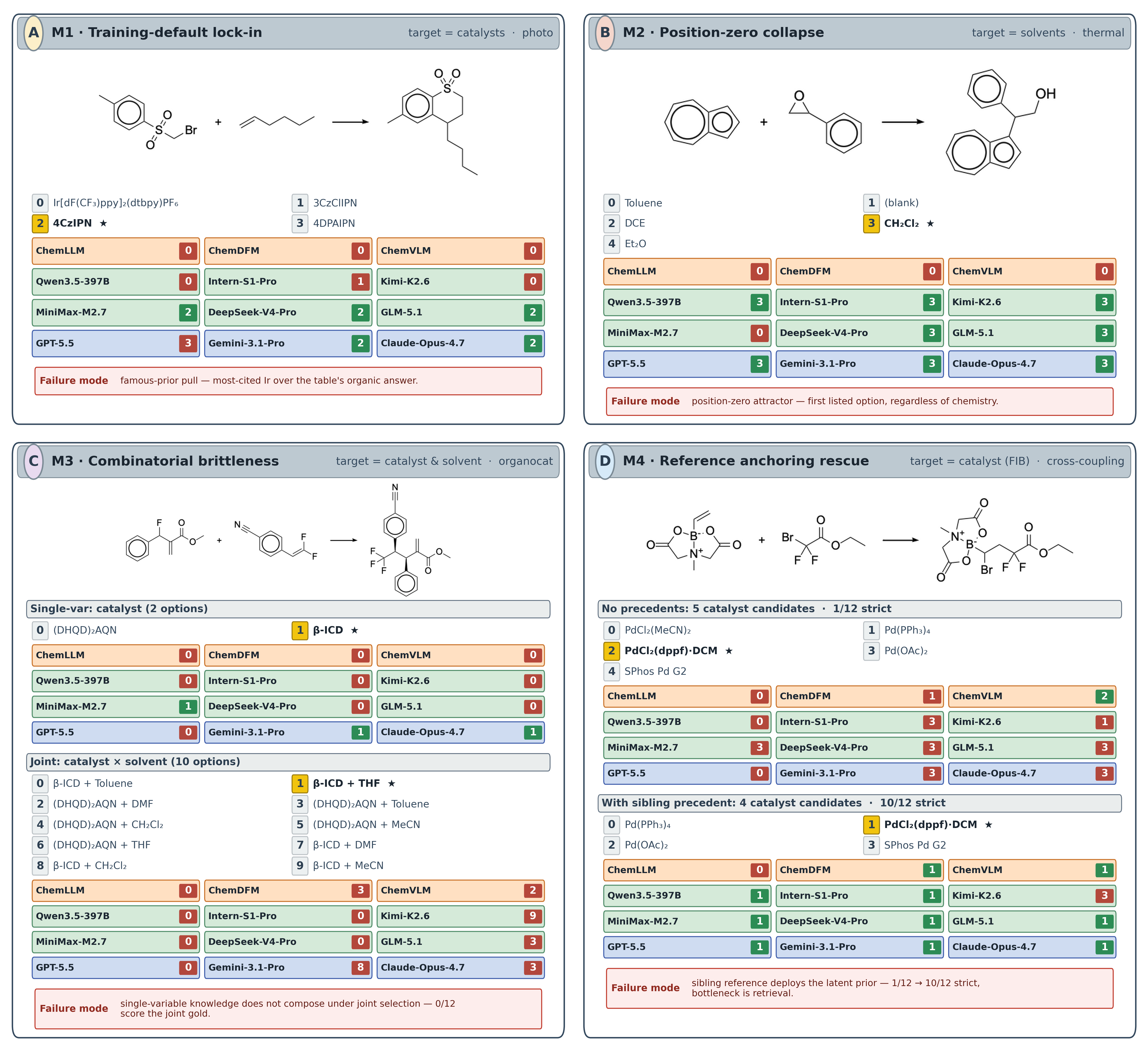}
\caption{\textbf{Per-model predictions on four illustrative cases.} Each card shows the target reaction, options (gold $\bigstar$), and the 12 models' picks (green $=$ correct, red $=$ wrong). M1/M2: \texttt{MC\_single} items; M3: paired \texttt{MC\_single}/\texttt{MC\_all} on the same parent reaction; M4: FIB without vs.\ with a sibling precedent.}
\label{fig:case_studies}
\end{figure}

Aggregate accuracy hides item-level behavior. We therefore selected four
single-reaction probes (Figure~\ref{fig:case_studies}) to illustrate
item-level behaviors that are not visible in Table~\ref{tab:all_heatmap} or
Table~\ref{tab:perkey}. These cases are qualitative examples, not an exhaustive
taxonomy, prevalence estimate, or causal validation of the underlying
mechanism.

\paragraph{(M1) Common-condition selection.}
On a photoredox optimisation (gold $=$ 4CzIPN at idx $=2$), $5/12$ models --- the Chemistry trio plus Qwen3.5 and Kimi --- select the widely used Ir[dF(CF$_3$)ppy]$_2$(dtbpy)PF$_6$ at idx $=0$. This pattern is consistent with a strong default prior, but the example alone cannot distinguish that explanation from other causes.

\paragraph{(M2) Index-zero selection.}
A BF$_3$-catalysed azulene rearrangement places non-default solvents at both idx $=0$ and idx $=3$; the Chemistry trio (plus MiniMax-M2.7) select idx $=0$ while several frontier models converge on the gold. Together with the corpus-level index-frequency diagnostic, this is consistent with position sensitivity; it does not establish chemistry-domain finetuning as its cause.

\paragraph{(M3) Combinatorial brittleness.}
The same parent reaction is asked as a 2-option catalyst question ($\beta$-ICD vs.\ (DHQD)$_2$AQN) and as a 10-option catalyst$\times$solvent question (gold $=$ $\beta$-ICD$+$THF). Three strong models recover the catalyst alone, but every scoring model fails the joint version --- single-variable recall does not compose into joint-condition accuracy. This brittleness echoes the aggregate pattern in Figure~\ref{fig:reference_mc}: precedents lift \texttt{MC\_single} Accuracy by $+13$--$18$\,pp across frontier models, yet on \texttt{MC\_all} the same precedents go neutral or negative --- a signal that disambiguates one axis becomes distractor noise once the model must couple two.

\paragraph{(M4) Precedent-anchored rescue.}
A Pd-catalysed cross-coupling (gold catalyst $=$ PdCl$_2$(dppf)$\cdot$DCM) jumps from $1/12$ to $10/12$ strict when a sibling precedent is supplied, illustrating that a relevant observation can substantially alter the decision. M3--M4 also leave aggregate traces in Table~\ref{tab:all_heatmap} and Figures~\ref{fig:reference_mc}--\ref{fig:reference_fib}; M1--M2 are illustrated here only at the item level.

\clearpage
\section*{NeurIPS Paper Checklist}

\begin{enumerate}

\item {\bf Claims}
    \item[] Question: Do the main claims made in the abstract and introduction accurately reflect the paper's contributions and scope?
    \item[] Answer: \answerYes{}
    \item[] Justification: The abstract and introduction claim (i) a real-data benchmark for reaction-condition optimization with paired precedents/no-precedents items, (ii) a continuous yield-and-selectivity scoring scheme, (iii) public release of the final human-reviewed benchmark test set and evaluation code, and (iv) a systematic evaluation across nine frontier and three Chemistry LLMs showing substantial headroom. Each claim is supported in the paper: (i) by Section~\ref{sec:question_construction} and Section~\ref{sec:refctrl}; (ii) by Section~\ref{sec:metrics} (Eq.~1); (iii) by Section~\ref{sec:pipeline} and Appendix~\ref{app:license}; and (iv) by the results in Section~\ref{sec:results}.

\item {\bf Limitations}
    \item[] Question: Does the paper discuss the limitations of the work performed by the authors?
    \item[] Answer: \answerYes{}
    \item[] Justification: Section~5 (\emph{Conclusion and Future Work}) acknowledges current scope and operational caveats through the refinements it commits to in subsequent releases: (a) coverage gaps in undersampled condition families (electrochemistry, photochemistry, flow chemistry, biocatalysis) and in non-English methodology journals; (b) the order-dependence of the multi-component condition comparator (e.g., reagent or solvent mixtures), which a planned order-invariant FIB comparator will address; (c) the single-number nature of the effective score, which the planned multi-objective scoring layer will extend; and (d) the gap exposed by case study M4 between possessing and deploying a relevant prior, which the planned retrieval-augmented evaluation setting targets. Periodic RxnOptBench refreshes drawn from newly published methodology papers further mitigate training-corpus contamination as the field evolves.

\item {\bf Theory assumptions and proofs}
    \item[] Question: For each theoretical result, does the paper provide the full set of assumptions and a complete (and correct) proof?
    \item[] Answer: \answerNA{}
    \item[] Justification: The paper makes no formal theoretical claims; it is an empirical benchmark paper.

\item {\bf Experimental result reproducibility}
    \item[] Question: Does the paper fully disclose all the information needed to reproduce the main experimental results of the paper to the extent that it affects the main claims and/or conclusions of the paper?
    \item[] Answer: \answerYes{}
    \item[] Justification: The four-stage extraction methodology is described in Section~\ref{sec:pipeline}; question construction in Section~\ref{sec:question_construction}; metrics with closed-form definitions in Section~\ref{sec:metrics} (Eq.~1); and the inference protocol in Section~\ref{sec:models}. The public release contains the final human-reviewed benchmark test set and deterministic evaluation code, allowing supplied predictions to be scored consistently. Extraction and benchmark-construction implementations and the archived model outputs used in the paper are not part of the public release; exact regeneration from raw PDFs or exact replay of hosted-model responses is not claimed.

\item {\bf Open access to data and code}
    \item[] Question: Does the paper provide open access to the data and code, with sufficient instructions to faithfully reproduce the main experimental results, as described in supplemental material?
    \item[] Answer: \answerYes{}
    \item[] Justification: The public assets are limited to the final human-reviewed benchmark test-set records and evaluation code, with instructions for formatting predictions and reproducing benchmark scores. The extraction pipeline, benchmark-construction code, review tooling, and archived model outputs used in the paper are not public. The test-set records will use a lightweight research-use agreement in response to the identified aggregation risk. Access terms, hosting, intended use, and worked examples are documented in Appendices~\ref{app:license} and~\ref{app:subtypes}.

\item {\bf Experimental setting/details}
    \item[] Question: Does the paper specify all the training and test details (e.g., data splits, hyperparameters, how they were chosen, type of optimizer, etc.) necessary to understand the results?
    \item[] Answer: \answerYes{}
    \item[] Justification: No model is trained in this work --- all 12 evaluated models are queried as-is. Inference details are reported in Section~\ref{sec:models} (single shared prompt template per question type, greedy or provider non-sampling mode, and a single sample per question). The 4{,}773 questions form a single test set; train/validation splits are not applicable.

\item {\bf Experiment statistical significance}
    \item[] Question: Does the paper report error bars suitably and correctly defined or other appropriate information about the statistical significance of the experiments?
    \item[] Answer: \answerNo{}
    \item[] Justification: Models are queried once per question with sampling disabled, which does not yield seed-level variance estimates. All models are evaluated on the same 4{,}773-item benchmark. The headline no-precedents subsets contain 1,039 \texttt{MC\_single}, 552 \texttt{MC\_all}, and 552 FIB questions; the remaining items form the matched precedents comparisons.

\item {\bf Experiments compute resources}
    \item[] Question: For each experiment, does the paper provide sufficient information on the computer resources (type of compute workers, memory, time of execution) needed to reproduce the experiments?
    \item[] Answer: \answerYes{}
    \item[] Justification: Open-weight models were served on local GPU nodes and proprietary models were queried through their public APIs. The evaluation consists of one non-sampling pass over the fixed 4{,}773-question test set; the model serving regime and pipeline components are described in Sections~\ref{sec:models} and~\ref{sec:pipeline}.

\item {\bf Code of ethics}
    \item[] Question: Does the research conducted in the paper conform, in every respect, with the NeurIPS Code of Ethics?
    \item[] Answer: \answerYes{}
    \item[] Justification: Source documents were manually obtained through open-access channels or university institutional subscriptions for noncommercial research; no publisher-site crawling was used and no source PDFs, images, or verbatim text are redistributed. Applicable publisher TDM and release terms will be verified before final release. No personal or sensitive data are included. Verification was performed by six paid interns in mainland China at the disclosed rates (Appendix~\ref{app:safe:ui}); no human-subjects research is involved.

\item {\bf Broader impacts}
    \item[] Question: Does the paper discuss both potential positive societal impacts and negative societal impacts of the work performed?
    \item[] Answer: \answerYes{}
    \item[] Justification: The benchmark supports evidence-based evaluation of LLM condition recommendation and reduces overclaiming from shallow proxies. Appendix~\ref{app:license} now also explains that machine-readable aggregation lowers the barrier to reuse and may enable unvalidated wet-lab recommendations. Mitigations include a research-use agreement, retained provenance, explicit expert-review and experimental-validation warnings, and a misuse/rights reporting channel.

\item {\bf Safeguards}
    \item[] Question: Does the paper describe safeguards that have been put in place for responsible release of data or models with a high risk for misuse (e.g., pretrained language models, image generators, or scraped datasets)?
    \item[] Answer: \answerYes{}
    \item[] Justification: Although no pretrained model or source publication is redistributed, structuring published reaction-condition records lowers access friction. Final records will therefore use a lightweight research-use agreement requiring source attribution, compliance with applicable law/source terms, and acknowledgement that outputs require qualified expert review and independent experimental validation (Appendix~\ref{app:license}).

\item {\bf Licenses for existing assets}
    \item[] Question: Are the creators or original owners of assets (e.g., code, data, models), used in the paper, properly credited and are the license and terms of use explicitly mentioned and respected?
    \item[] Answer: \answerYes{}
    \item[] Justification: All 232 source papers are cited by DOI in the released metadata. External tooling --- DECIMER-Segmentation~\citep{rajan2021decimerseg}, DECIMER~\citep{rajan2023decimer}, GTR-VL~\citep{wang2025gtr}, RDKit~\citep{landrum2024rdkit}, PubChem~\citep{kim2025pubchem} --- is cited in Section~\ref{sec:pipeline}; their open-source licenses are respected and listed in Appendix~\ref{app:license}. Evaluated models are credited in the references and used under their published API terms or open-weight licenses (Section~\ref{sec:models}).

\item {\bf New assets}
    \item[] Question: Are new assets introduced in the paper well documented and is the documentation provided alongside the assets?
    \item[] Answer: \answerYes{}
    \item[] Justification: A datasheet covering motivation, composition, collection, preprocessing, uses, distribution, and maintenance will accompany the camera-ready release; license, hosting, and intended use are detailed in Appendix~\ref{app:license}. Worked question examples (Appendix~\ref{app:subtypes}) and the human-verification protocol (Appendix~\ref{app:safe:ui}) document the test set and its curation methodology.

\item {\bf Crowdsourcing and research with human subjects}
    \item[] Question: For crowdsourcing experiments and research with human subjects, does the paper include the full text of instructions given to participants and screenshots, if applicable, as well as details about compensation (if any)?
    \item[] Answer: \answerYes{}
    \item[] Justification: The verification stage employed four master's-level and two doctoral-level paid interns based in mainland China, compensated at RMB~300 and RMB~400 per working day, respectively. They were project collaborators rather than anonymous crowdworkers. The interface, decisions (accept / reject / edit), workflow, region, and rates are documented in Appendix~\ref{app:safe:ui}.

\item {\bf Institutional review board (IRB) approvals or equivalent for research with human subjects}
    \item[] Question: Does the paper describe potential risks incurred by study participants, whether such risks were disclosed to the subjects, and whether Institutional Review Board (IRB) approvals (or an equivalent approval/review based on the requirements of your country or institution) were obtained?
    \item[] Answer: \answerNA{}
    \item[] Justification: Annotators review chemistry tables from public peer-reviewed papers; no human-subjects data is collected, no behavioral or biomedical experimentation is performed on the annotators, and no personal information is processed. IRB review is therefore not applicable under our institution's policy.

\item {\bf Declaration of LLM usage}
    \item[] Question: Does the paper describe the usage of LLMs if it is an important, original, or non-standard component of the core methods in this research?
    \item[] Answer: \answerYes{}
    \item[] Justification: VLMs are a core component of the extraction pipeline (page selection, table understanding, label-to-structure binding, per-row verification) and their usage, prompting constraints, and human-review gating are described in detail in Section~\ref{sec:pipeline}. LLM/VLM models evaluated as benchmark subjects are listed in Section~\ref{sec:models}. LLMs were not used to write the scientific content of the paper beyond standard editorial assistance.

\end{enumerate}

\end{document}